\catcode`\@=11
\newif\if@fewtab\@fewtabtrue

{\count255=\time\divide\count255 by 60
\xdef\hourmin{\number\count255}
\multiply\count255 by-60\advance\count255 by\time
\xdef\hourmin{\hourmin:\ifnum\count255<10 0\fi\the\count255}}
\def\ps@draft{\let\@mkboth\@gobbletwo
    \def\@oddhead{}
    \def\@oddfoot
       {\hbox to 7 cm{$\scriptstyle Draft\ version:\ \draftdate$
       \hfil}\hskip -7cm\hfil\rm\thepage \hfil}
    \def\@evenhead{}\let\@evenfoot\@oddfoot}

\def\ceqno{\global\@fewtabfalse
    \ifcase\@eqcnt \def\@tempa{& & &}\or \def\@tempa{& &}
      \or \def\@tempa{&}
      \or\def\@tempa{}\fi\@tempa
{\rm(\theequation)}}

\def\aeqno#1{\global\@fewtabfalse
    \ifcase\@eqcnt \def\@tempa{& & &}\or \def\@tempa{& &}
      \or \def\@tempa{&}
      \or\def\@tempa{}\fi\@tempa
{\rm(\theequation,#1)}}

\def\label#1{\ifnum\draftcontrol=1
 \global\def\draftnote{$\scriptstyle #1$}\fi
 \@bsphack\if@filesw {\let\thepage\relax
   \def\protect{\noexpand\noexpand\noexpand}%
\xdef\@gtempa{\write\@auxout{\string
      \newlabel{#1}{{\@currentlabel}{\thepage}}}}}\@gtempa
   \if@nobreak \ifvmode\nobreak\fi\fi\fi
  \@esphack}

\def\alabel#1#2{\label{#1}\global\@fewtabfalse
    \ifcase\@eqcnt \def\@tempa{& & &}\or \def\@tempa{& &}
      \or \def\@tempa{&}
      \or\def\@tempa{}\fi\@tempa
{\hbox to 3cm{\phantom{\rm(\theequation,#2)}
\draftnote \hfil}\hskip -3cm {\rm(\theequation,#2)}}}

\def\clabel#1{\label{#1}\global\@fewtabfalse
    \ifcase\@eqcnt \def\@tempa{& & &}\or \def\@tempa{& &}
      \or \def\@tempa{&}
      \or\def\@tempa{}\fi\@tempa
{\hbox to 3cm{\phantom{\rm(\theequation)}
\draftnote \hfil}\hskip -3cm{\rm(\theequation)}}}

\def\eqnarray{\def\draftnote{{}}\global\@fewtabtrue
\stepcounter{equation}\let\@currentlabel=\theequation
\global\@eqnswtrue
\global\@eqcnt\z@\tabskip\@centering\let\\=\@eqncr
$$\halign to \displaywidth\bgroup\@eqnsel\hskip\@centering\@eqcnt\z@
  $\displaystyle\tabskip\z@{##}$&\global\@eqcnt\@ne
  \hskip 1\arraycolsep \hfil${##}$\hfil
  &\global\@eqcnt\tw@ \hskip 1\arraycolsep
$\displaystyle\tabskip\z@{##}$
\hfil  \tabskip\@centering&\global\@eqcnt\thr@@\llap{##}\tabskip\z@
\cr}

\def\endeqnarray{\@@eqncr\egroup
      \global\advance\c@equation\m@ne$$\global\@ignoretrue}

\def\@eqnnum{\hbox to 3cm{\phantom{\rm(\theequation)} \draftnote
                         \hfil}\hskip -3cm {\rm(\theequation)}}

\def\@@eqncr{\let\@tempa\relax
    \ifcase\@eqcnt \def\@tempa{& & &}\or \def\@tempa{& &}
      \or \def\@tempa{&}
      \or\def\@tempa{}
\fi\@tempa
\if@eqnsw
\if@fewtab\@eqnnum\fi
\stepcounter{equation}\fi\global
\@eqnswtrue\global\@eqcnt\z@\global\@fewtabtrue\cr}

\def\draftcite#1{\ifnum\draftcontrol=1#1\else{}\fi}

\def\@lbibitem[#1]#2{\item{}\hskip -3cm \hbox to 2cm
{\hfil$\scriptstyle\draftcite{#2}$}\hskip
1cm[\@biblabel{#1}]\if@filesw
     {\def\protect##1{\string ##1\space}\immediate
      \write\@auxout{\string\bibcite{#2}{#1}}}\fi\ignorespaces}

\def\@bibitem#1{\item\hskip -3cm \hbox to 2cm
{\hfil $\scriptstyle\draftcite{#1}$}\hskip 1cm
\if@filesw \immediate\write\@auxout
       {\string\bibcite{#1}{\the\value{\@listctr}}}\fi\ignorespaces}

     \def\nappendix#1{\vskip 1cm\no{\bf Appendix
         #1}\def\thesection{#1} \setcounter{equation}{0}}

\font\tendl=msbm10  scaled \magstep1
\font\sevendl=msbm7 scaled \magstep1
\font\fivedl=msbm5 scaled \magstep1
\font\tengl=eufm10  scaled \magstep1
\font\sevengl=eufm7 scaled \magstep1
\font\fivegl=eufm5 scaled \magstep1

\newfam\dlfam \def\dl{\fam\dlfam\tendl} 
\textfont\dlfam=\tendl \scriptfont\dlfam=\sevendl
\scriptscriptfont\dlfam=\fivedl
\newfam\glfam  
\textfont\glfam=\tengl \scriptfont\glfam=\sevengl
\scriptscriptfont\glfam=\fivegl

\def\draftdate{\number\month/\number\day/\number\year\ \ \ \hourmin }

\global\def\draftcontrol{0}
\catcode`\@=12
\def\tilde{\widetilde}
\def\hat{\widehat}

\documentstyle[12pt,epsf]{report}

\def\theequation{{\thesection.\arabic{equation}}}

\newcommand{\be}{\begin{eqnarray}}
\newcommand{\en}{\end{eqnarray}\vs 0.5 cm}

\newcommand{\no}{\noindent}
\newcommand{\vs}{\vskip}

\newcommand{\NR}{{{\dl R}}}
\newcommand{\NZ}{{{\dl Z}}}
\newcommand{\NN}{{{\dl N}}}

\newcommand{\qq}{\begin{eqnarray}}

\newcommand{\qqq}{\end{eqnarray}}

\newcommand{\CD}{{\cal D}}

\newcommand{\CF}{{\cal F}}

\newcommand{\CH}{{\cal H}}

\newcommand{\CL}{{\cal L}}

\newcommand{\CN}{{\cal N}}

\newcommand{\CW}{{\cal W}}
\newcommand{\CX}{{\cal X}}

\begin{document}

\thispagestyle{empty} 
~\\
\vspace{15mm}
\begin{center}
{\huge \bf Horizon Branes and Chiral Strings}\\
\vspace{20mm}
{\Large \bf Nuno Reis}\\
\vspace{10mm}
{\Large
Theoretical Physics, Department of Physics,\\
University of Oxford,\\
\vskip3mm
1 Keble Road, Oxford OX1 3NP, United Kingdom
}
\end{center}
\vspace{30mm}
\begin{center} 
{\Large\font\oxcrest=oxcrest40 \oxcrest\char'01}
\end{center} 
\vspace{40mm}
\begin{center}
{\large Thesis submitted in partial fulfillment of the requirements\\
for the degree of Master of Science by Research\\
in the University of Oxford\\ 
\vskip3mm
- Trinity Term 2002 -}
\end{center}



\newpage
\begin{center}
{\huge \bf Abstract}
\end{center}
\vspace{5mm}
\hskip0.5cm String Theory is known to be one possible model to unify all the known forces 
of the Nature by a universal concept, that of strings. All fundamental particles, including 
gravitons with a given energy, are supposed to be oscillating states of tensive 
open or closed strings. The string concept introduces some non-locality in the 
gravitational interactions, making the self-interacting Feynman diagrams finite order 
by order. The discovery of $D$-branes has created a new situation in string theory,  
as those objects are non-perturbative solutions dual to solitonic charged objects 
describing black $p$-branes, solutions of the supergravity equations. 
Since then many occurences of dualities have been found in String Theory where all
five different type of vacua are related by M-theory. On the other hand, a well defined 
quantum gravity theory should statistically count the number of quantum states giving 
rise to the Bekenstein-Hawking entropy of ${\it any}$ black hole horizon. A partial 
success has been achieved in the particular case when the near horizon geometry is 
that of Anti-de Sitter, giving rise to the general AdS/CFT holographic principle. 

In this thesis I present a new type of brane - $H$-brane - where the role of time 
in String Theory is considered as a primary concept in the search for still 
unknown black hole physics. We start to study the physics of charged open strings immersed in a critical electric field. Using the lightcone gauge, we find that open strings are naturally described by a worldsheet with null boundaries that define the $H$-brane.  Using the basic tools of boundary conformal field theory, 
we describe the $H$-branes both in the open and closed string channels and examine how 
they fit naturally in the known $D$-brane moduli space. In particular we compute their Ishibashi states and quantize the system using the first order formalism. We find that the geometry associated to target null coordinates is non-commutative. This is a possible way to solve the information loss paradox.

We conjecture that {\it any quantum horizon - black hole, cosmological, etc. - is 
phenomenologically described in time-dependent String Theory by a chiral and 
non-normalized squeezed state. }
\newpage
\begin{center}
{\huge \bf Acknowledgments}
\end{center}
\vskip0.5cm
\hskip0.5cm Firstly I would like to thank my supervisor Ian Kogan for sharing his ideas on black hole area quantization using chiral closed strings and for his encouragement to find a complementary picture to the well established $AdS$$/CFT$ correspondance to explain statistically 
the Bekenstein-Hawking entropy of any black hole horizon using string theory, which have 
lead us to the discovery of $H$-branes.\vskip0.3cm

Further I would like to thank Krzysztof Gawedzki for very useful remarks on the formulation of our $H$-branes and for reading the manuscript. Also to my examinors, Dr. Jos\'e Figueroa-O'Farrill and  Dr. John F. Wheater, for corrections and remarks on my thesis.   \vskip0.3cm

To the people of Oxford for the pleasant times we have spent together, in particular to Mario Santos for his friendsheap. To Shinsuke Kawai for stimulating and very useful discussions. To my friends Ana, Alex and Rui and in special to my Mother for supporting me during the difficult times.\vskip0.3cm

To Marta for the last years and her comprehension over the time I have spent in Oxford.\vskip0.3cm

I acknowledge Linacre College and the Theoretical Physics Group for accepting me and
for creating the opportunity to develop my work.\vskip0.3cm

The thesis was supported by the grant PRAXIS XXI/BD/18138/98 from FCT (Portugal).

\newpage
~\\
\vspace{60mm}
\begin{center}
{\huge \bf Publication}
\end{center}
\vskip0.5cm
Ian Kogan, Nuno Reis: {\it H-Branes and Chiral Strings}. Int.J.Mod.Phys. {\bf A16} (2001), 4567-4590; arXiv:hep-th/0107163.
\newpage
\tableofcontents




%
%




%
%
%
\chapter{Introduction}
\vskip0.3cm
\hskip0.5cm 
The discovery of $D_p$-branes by Polchinski provided a non-perturbative 
string theory description
of the supergravity black $p$-brane solutions carrying RR charges. Since then 
many dualities in String Theory have been found where non-perturbative effects 
become an accessible problem after a strong-weak coupling duality. String Theory 
has taught us that $D$-branes provide a useful tool to obtain a microscopic (dual) 
picture of the Bekenstein-Hawking entropy of an extremal charged black hole.
Near the horizon, such extremal black holes are solutions of supergravity 
on $AdS$ which, by the holographic principle, are dual to the  $\CN=4$ Yang-Mills 
theory at the fixed point. One counter example where the use of duality does not permit us to
solve the problem is  the relative motion of a $D$-brane which is dual 
to a free open string, with endpoints carrying an electric charge $e$, immersed in 
a constant electric field.
The unknown behavior of the system when the electric field approaches the critical 
limit $E\rightarrow E_{crit} = (2\pi\alpha' e)^{-1}$ is dual to the unknown behavior 
of a D-brane moving with velocity close to the speed of light $V\rightarrow  V_{crit}=1$. 
What is missing in the known $D$-brane moduli space is some kind of an infinitely 
boosted brane solution.

In \cite{KR} the existence of a new type of brane in the time 
dependent string theory, namely the nullbranes, was postulated. The nullbranes
were introduced in the literature to provide a supplementary tool to study unknown 
properties of black holes for more general situations than that of the extremal 
charged regime. In the closed string channel, our nullbranes 
are described by chiral and non-normalized squeezed Ishibashi states, the properties 
that are believed to be phenomenologically associated to quantum event horizons - 
 black holes, cosmological (de-Sitter), etc. For this reason our nullbrane was renamed 
as a {\bf Horizon-brane}, or $H$-brane for short \footnote{Let us note that nevertheless there 
is some information about null structure in the letter $H$ too: Russian $H$ = Latin $N$.}. 

This thesis is organized as follows. In Sect. 2.1 we start to review some mathematical techniques of quantization of $d$-dimensional 
systems, as applied to $d=2$ field theory at a renormalization group fixed point
which in Sect. 2.2 and Sect. 2.3 we see to provide the worldsheet description of a closed or open string respectively.
In Sect. 2.4 we focus on the case of boundary conformal field theory (BCFT), useful for the treatment 
of $D$-branes in flat backgrounds. Our aim is to study the $D$-brane behavior under 
an infinite boost. In Sect. 2.5 we start to calculate the correlation functions for a free string 
perturbed by an external electromagnetic field and in Sect. 2.6  we review the general 
mathematical treatment of boundary perturbations by self-dual fields, where the Dissipative Hofstadter Model is one particular example.

 In Sect. 3.1 we describe how boosted $D$-branes in the Minkowskian 
spacetime are related by $T$-duality to the charged open strings immersed in a constant 
electric field and explain why there is an upper limit for the electric field 
at which the theory breaks down.

 From the results in the previous sections on the two-point 
correlation functions, we consider two distinct cases of charged open 
strings, one with a magnetic background (Sect. 3.2) and the other one with an electric 
background (Sect. 3.3). We show how the magnetic case leads to non-commutativity 
of space coordinates and gives some hints about the spacetime geometry in the electric 
case. Sect. 3.4 analyzes the tensionless strings for which all points travel at the speed 
of light. Such strings are described by the Schild action whose quantization 
leads to non-commutativity of the spacetime coordinates of string endpoints. 

In Sect. 4.1 we give the technical definition of an $H$-brane in terms of boundary conditions and see how they fit naturally in a Bosonic String Theory in the lightcone gauge. We proceed to their quantization in Sect. 4.2 using the first order formalism, where there is no room for guesswork. In Sect. 4.3 we give some hints on the supersymmetric extension to our $H$-branes. Sect. 4.4 shows that there is enough structure in the null directions so that we may neglect one chiral sector of the closed string without introducing singularities into the system. In Sect. 4.5 we show that chiral closed strings are coupled naturally to $H$-branes and calculate the Ishibashi states by the first order formalism.

In Sect. 5 we give some hints on our motivations to describe quantum black hole horizons by H-branes.

Conclusions give a brief summary of what was achieved in the thesis and discuss open questions to be addressed in the future. More technical calculations referred to in the main text have been collected in Appendices.


\newpage
\chapter{D-branes and Conformal Field Theory}
\section{The first-order formalism}

\hskip0.5cm To find the symplectic structure 
of the phase space of a physical system and proceed with the Dirac quantization, we use the first order formalism where there is no room for guesswork 
\cite{gawedzki}. The basic object is a $d$-form $\alpha$ on a bundle
over the $d$-dimensional 
spacetime $\Sigma$ through which the first-order action is
expressed. 
In our case it is $d=2$ worldsheet with the string target coordinates 
considered as fields $\Phi$ depending on two parameters $\tau$ and
$\sigma$. The 
original worldsheet action is of the form
\qq
S=\int_{\Sigma}\CL(x^a, \phi^\mu, \partial_a\phi^\mu)
\qqq
where $\mu$ run over the target coordinates $0,...,D-1$ and $a$ over the two-dimensional 
worldsheet space $\Sigma$ that might or might not have a boundary $\partial\Sigma$ depending on whether the string is open or closed respectively. In general, such an action leads to
variational equations with second order derivatives.  To obtain the first-order formalism,
one defines  
\qq
\xi_a^\mu=\partial_a\phi^\mu\,\,\,,\hskip1.5cm \Pi^a_\mu=\frac{\partial \CL}{\partial \xi_a^\mu}
\qqq
and introduces the 2-form
\qq
\alpha=\CL dx^0\wedge dx^1\,+\,\Pi^0_\mu(d\phi^\mu-\xi_0^\mu dx^0)
\wedge dx^1+\Pi^1_\mu dx^0\wedge(d\phi^\mu-\xi_1^\mu dx^1)
\qqq
on the space with coordinates $(x^a,\phi^\mu,\xi^\mu_a)$ which forms a bundle $P$
over $\Sigma$. The fields $\Phi=(\phi^\mu, \xi_a^\mu)$ are geometrically interpreted 
as sections of $P$. The first-order action is given by 
\qq
S'=\int_{\Sigma} \Phi^\ast\alpha
\qqq
and it coincides with the original one if $\xi^\mu_a=\partial_a\phi^\mu$.
The variational equations $\delta S'(\Phi)=0$ take the geometric form
\qq
\Phi^\ast(i(\delta \Phi) d\alpha)=0\hskip1.8cm {\rm on}\hskip0.7cm \Sigma\nonumber\\
\Phi^\ast(i(\delta \Phi)\alpha)=0\hskip1.5cm {\rm along}\hskip0.5cm \partial\Sigma
\qqq
for any vector field $\delta\Phi$ tangent to $P$ and respecting the boundary condition
when present (such vector fields describe infinitesimal variations of $\Phi$). 
The bulk equations reduce to the standard Euler-Lagrange equations plus 
the conditions $\xi_a^\mu=\partial_a\phi^\mu$. The boundary equations may 
additionally restrict the behavior of the solutions on the boundary. 
The space of classical solutions 
carries a closed 2-form
\qq
\Omega(\delta_1\Phi,\delta_2 \Phi)=\int_{\Sigma_t}\Phi^\ast(i(\delta_2 \Phi) i(\delta_1 \Phi) d\alpha)\,-\,\int_{\partial\Sigma_t}\Phi^\ast(i(\delta_2 \Phi) i(\delta_1 \Phi)\alpha)
\qqq
where $\delta_i\Phi$ are  vectors tangent to the space of the classical solutions 
and $\Sigma_t$ is a slice of the worldsheet, usually the line of constant time where 
the Cauchy data may be defined. If the form $\Omega$ is non-degenerate, then it
provides the space of classical solutions with a symplectic structure. Otherwise
one should identify the solutions that are connected by one-parameter families
tangent to the degeneration directions. $\Omega$ descends then to a symplectic 
form on the quotient space $\tilde P$ that forms the phase space of the theory.
The symplectic structure leads to the Hamiltonian vector fields $X_{\CF}$ 
corresponding to functions $\CF$ of the phase space, such that $d\CF=i_{X_{\CF}}\Omega$ 
and the Poisson bracket $\{\CF,\CF'\}=X_{\CF}(\CF')$.


\section{Closed strings and Conformal Field Theory}

\hskip0.5cm There is a well known  relation between closed strings in curved spacetimes $M$ 
and the two-dimensional sigma model defined on closed $2d$ worldsheets 
$\Sigma$ with the spacetime $M$ as the target. Let $X^\mu$ denote the coordinates
of $M$ and $G_{\mu\nu}$ and $R$ the metric tensor and the Ricci scalar of $M$, respectively.
We also turn on the Kalb-Ramond $B_{\mu\nu}$ field tensor. The closed string coupling 
constant $g_c$ is given by the dilaton field as $g_c\propto e^\Phi$. 
The closed string immersed in a curved spacetime is described by specifying the
dependence of the coordinates $X^\mu$ on the worldsheet point. 
The string tension $T$ is given by the Regge slope as $T=1/(2\pi\alpha')$.
From the non-linear sigma model with the action
\qq
S_\sigma=\frac{1}{4\pi\alpha'}\int d\tau d\sigma \gamma^{1/2}\left[(\gamma^{ab}G_{\mu\nu}(X)+i\epsilon^{ab}B_{\mu\nu}(X))\partial_aX^\mu\partial_bX^\nu\,+\,\alpha'R\Phi(X)\right]
\label{nons}
\qqq
one calculates the trace of the energy-momentum tensor and from that the renormalization group $\beta$-functions describing the scale dependence
of $G_{\mu\nu}$, $B_{\mu\nu}$ and $\Phi$. The string equations of motion are defined 
by imposing the fixed point condition on the sigma model ($\beta=0$), as required 
by the Weyl invariance, i.e. closed strings are described by $2d$ Conformal Field Theory 
(CFT) on the worldsheet. Performing perturbation theory by expanding the fields around 
flat solutions, we may explore the low-energy regime for which the radius of curvature is much larger then the characteristic string length scale $\alpha'^{1/2}$. In the 
target space, the equations of motion are described by the low-energy action
of an effective field theory, as we have ignored the internal structure of the string.

On the other hand, we can associate order by order perturbative terms in the sigma model 
to vertex operators in the target space. The scattering of $n$ particles with each of them 
carrying momentum $k_i$ is given by the $S$-matrix with entries
\qq
S_{j_1,...,j_n}(k_1,...,k_n)=\hskip8.5cm\nonumber\\
\hskip1.0cm\sum \int\frac{\CD X \,\CD \gamma}{Vol_{diff\times Weyl}}\,e^{-S_\sigma-\lambda\chi} 
\prod_{i=1}^n\int V_i(k_i;\tau,\sigma){\gamma^{1/2}(\tau,\sigma)}d\tau d\sigma
\label{sm}
\qqq
where $V_i$ is the vertex operator inserted at position $(\tau,\sigma)$ on the worldsheet 
and the sum is over all compact 2d topologies. They are characterized by the Euler number 
$\chi$ and the measure is normalized by the volume of the diffeomorphism and Weyl symmetry 
groups for each compact worldsheet. $S_\sigma$ is the action of the sigma model for a flat spacetime. We obtain the vertex operators by, for example, expanding about the flat spacetime, $G_{\mu\nu}=\eta_{\mu\nu}+\chi_{\mu\nu}(X)$, in (\ref{nons}), 
substituting into the Polyakov path-integral and comparing order by order with (\ref{sm}). 
The result is that a curved spacetime is in fact a coherent background of gravitons 
whose vertex operators are of the form
\qq
V_G\propto -4\pi g_c\,\partial_aX^\mu\partial_bX^\nu e^{ik\cdot X}\gamma^{ab}\,\chi_{\mu\nu}
\label{gvo}
\qqq 
The CFT worldsheet description of string theory comes as one of the fundamental ideas of string theory to treat spacetime as a derived concept rather than as part of the input data \cite{schomerus}.

\vskip0.3cm 
To import the known mathematical techniques of CFT from statistical physics, let us 
restrict ourselves to a closed superstring model in a Minkowski spacetime with vanishing 
Kalb-Ramond and dilaton fields with the action  
\qq
S=\frac{1}{4\pi\alpha'}\int d\tau d\sigma \gamma^{ab}\partial_a X_\mu \partial_b X^\mu + \frac{i}{2\pi}\int d\tau d\sigma(\Psi^\mu\partial_-\Psi_\mu\,+\,\tilde\Psi^\mu\partial_+\tilde\Psi_\mu)
\qqq
with $\partial_\pm=(\partial_\tau\pm\partial_\sigma)/2$
and $\Psi^\mu$ are $2d$ Majorana fields. The general solution of the equations of motion
is given by splitting the fields into left- and right-chiral movers
\qq
X^\mu(\tau,\sigma)=X^\mu(x^+)+\tilde X^\mu(x^-) \hskip1.0cm \Psi^\mu(\tau,\sigma)=\left( \begin{array}{c}\Psi^\mu(x^-)\\ \tilde\Psi^\mu(x^+)\end{array}\right)
\qqq
with $x^\pm=\tau\pm\sigma$.
The action must be invariant under periodicity $\sigma\sim\sigma+2\pi$ which allows us to impose the conditions 
\qq
X^\mu(\tau,\sigma+2\pi)=X^\mu(\tau,\sigma)\nonumber\\ 
\Psi^\mu(\tau,\sigma+2\pi)=\exp^{(2\pi i\nu)}\Psi^\mu(\tau,\sigma)\nonumber\\ 
\tilde\Psi^\mu(\tau,\sigma+2\pi)=\exp^{(-2\pi i\tilde\nu)}\tilde\Psi(\tau,\sigma)
\qqq
where $\nu$ and $\tilde\nu$ can take the values $0$ or $1/2$, depending on whether
we choose the Ramond (R) or the Neveu-Schwarz (NS) sector, respectively. Note that 
as we have left- and right-chiral sectors, there are four $(\nu,\tilde\nu)$ possible 
types of closed superstrings.

The field Fourier expansions respecting the above periodicity conditions are 
\qq
X^\mu(\tau,\sigma)\,=\,\hat q^\mu\,+\,\hat\alpha_0^\mu x^-+\tilde\alpha_0^\mu x^+\,+\,i\alpha'^{1/2}\sum_{n\neq 0}(\frac{\alpha_n^\mu}{n}e^{-inx^-}\,+\,\frac{\tilde\alpha_n^\mu}{n}e^{-inx^+})\nonumber\\ \\
\Psi^\mu(\tau,\sigma)\,=\,\sum_{r\in\NZ+\nu}\psi^\mu_r e^{-ir x^-}\label{fX}\hskip0.45cm,\hskip0.65cm
\tilde\Psi^\mu(\tau,\sigma)\,=\,\sum_{r\in\NZ+\tilde\nu}\tilde\psi^\mu_r e^{-irx^+}\hskip0,6cm\nonumber
\qqq
with mode-oscillator elements whose Poisson brackets may be canonically quantized.

From the energy momentum tensor 
\qq
T_{++}\,=\,\frac{1}{\alpha'}\partial_+ X^\mu \partial_+X_\mu\,+
\,\frac{i}{2}\Psi^\mu\partial_+\Psi_\mu\nonumber\\
T_{--}\,=\,\frac{1}{\alpha'}\partial_-\tilde X^\mu\partial_-
\tilde X_\mu\,+\,\frac{i}{2}\tilde\Psi^\mu\partial_-\tilde\Psi_\mu
\qqq
and the supercurrents
\qq
J_+\,=\,\Psi^\mu\partial_+X_\mu\hskip0.3cm,\hskip0,6cm
J_-\,=\,\tilde\Psi^\mu\partial_-\tilde X_\mu
\qqq
we define their Fourier components by the following expressions
\qq
L_m=\frac{1}{2\pi}\int_0^{2\pi} d\sigma e^{imx^-}T_{--}\hskip0.3cm,\hskip0,6cm
\tilde L_m=\frac{1}{2\pi}\int_0^{2\pi} d\sigma e^{imx^+}T_{++}\nonumber\\
G_r=\frac{1}{2\pi}\int_0^{2\pi}d\sigma e^{irx^-}J_-\hskip0.3cm,\hskip0,6cm
\tilde G_r=\frac{1}{2\pi}\int_0^{2\pi}d\sigma e^{irx^+}J_+
\qqq
for $n\neq 0$ and $r \in \NZ+(\nu \,,\tilde \nu)$ as appropriate.

The canonical commutation relations for the holomorphic bosonic coordinates
give the relations on the mode oscillators
\qq
\left[\alpha_m^\mu,\alpha_n^\nu\right]=m\delta_{m+n,0}\eta^{\mu \nu}\hskip0,3cm,\hskip0,6cm\left[\hat x^\mu,\hat p^\nu\right]= i\eta^{\mu\nu}\hskip0,1cm.
\qqq
For the fermionic fields we obtain the canonical anti-commutation relations 
\qq
\{\psi^\mu_r,\psi^\nu_s\}=\eta^{\mu\nu}\delta_{r+s,0}
\qqq
with similar expressions on the anti-holomorphic fields. 
The Laurent modes for the energy-momentum tensor and for the supercurrent are given respectively by the expressions
\qq
L_m=\frac{1}{2}\sum_{n\in\NZ}:\alpha^\mu_{m-n}\alpha_{\mu n}:+\frac{1}{4}\sum_{r\in \NZ+\nu}(2r-m):\psi^\mu_{m-r}\psi_{\mu r}:+a\delta_{m,0}\nonumber\\
G_r=\sum_{n\in\NZ}\alpha^\mu_n\psi_{\mu r-n}\hskip7,8cm
\qqq
where $:\,\,:$ denotes creation-annihilation normal ordering and $a$ the normal ordering 
constant with the value depending on whether we are in the R-sector ($a=1/2$) or the 
NS-sector ($a=0$). The super Virasoro algebra follows 
\qq
\left[L_m,L_n\right]=(m-n)L_{m+n}+\frac{c}{12}m(m^2-m)\delta_{m+n,0}\nonumber\\
\{G_r,G_s\}=2L_{r+s}+\frac{c}{12}(4r^2-1)\delta_{r+s,0}\hskip2,0cm\\
\left[L_m,G_r\right]=\frac{m-2r}{2}G_{m+r}\hskip4,2cm\nonumber
\qqq
with similar expressions for the antiholomorphic sector. The conditions that a physical state associated to a closed string must obey are
\qq
L^{(C)}_m|\Psi_{phys}\rangle = 0\hskip1,3cm m>0\hskip0,3cm \\
L^{(C)}_0|\Psi_{phys}\rangle = (1/2-a) |\Psi_{phys}\rangle\nonumber\\
G^{(C)}_r|\Psi_{phys}\rangle = 0\hskip1,3cm r\geq 0\hskip0,4cm\nonumber
\qqq
with the same conditions involving the anti-holomorphic sector, e.g. $\tilde L^{(C)}_m|\Psi_{phys}\rangle = 0\,\,\,\,\,\forall m>0$.


\section{Open strings and Boundary CFT}

\hskip0,5cm We have seen in the last section how closed worldsheet diagrams are associated 
to closed string 
vacuum amplitudes. We shall now consider the diagrams with worldsheets with boundary 
associated to open string. The simplest case is obtained by restricting to the strip diagram with $-\infty<\tau<\infty$ and $0<\sigma<\pi$. 
By the map $(z,\bar z)=(e^{i\sigma -\tau},e^{-i\sigma-\tau})$ the strip is mapped to 
the upper half complex plane $\CH^+_0=\{z\,|\,\Im(z)\geq 0\}$ with boundaries $\sigma=0,\pi$ mapped to the real line $z=\bar z$. The holomorphic and anti-holomorphic sectors are reflected at the boundary, with oscillator modes related by a gluing map $\Omega$ 
\qq
\alpha^\mu_n+\Omega\tilde\alpha^\mu_{-n}=0\nonumber\\
\psi^\mu_r+\Omega\tilde\psi^\mu_{-n}=0
\qqq
at $z=\bar z$. Note that a creation oscillator mode is coupled to an annihilation mode.
The condition for no energy flow crossing the boundary is given by 
\qq
T(z)=\tilde T(\bar z)\hskip1.0cm
J(z)=\tilde J(\bar z)
\qqq
and it should be universal, i.e., it must not depend on the gluing map\,\footnote{We will 
see later that in fact we can relax such condition after we choose different type 
of worldsheet boundaries in time-dependent string theory}. Even if far away from the boundary 
the theory behaves as a usual CFT, the above super-Virasoro fields defined only 
at the upper half-plane, are not sufficient to give two commuting super-Virasoro algebras. 
Nevertheless we can construct {\it one} chiral algebra after we define the open super 
Virasoro operators by the method of images
\qq
L_n^{O}:=\frac{1}{2\pi i}\int z^{n+1} T(z)dz\,-\,\frac{1}{2\pi i}\int 
\bar z^{n+1}\tilde T(\bar z)d\bar z\nonumber\\
G_r^{O}:=\frac{1}{2\pi i}\int z^{n+1/2} J(z)dz\,-\,\frac{1}{2\pi i}\int 
\bar z^{n+1/2}\tilde J(\bar z)d\bar z
\qqq
where the integrals are over a semi-circle in $\CH^+_0$ and the label $(O)$ specifies 
that we are in the open channel.
This means that the coupled holomorphic and anti-holomorphic sectors of the boundary 
CFT defined in the upper half-plane are replaced by a holomorphic sector of a chiral CFT 
defined on the whole plane. The physical states must obey the conditions
\qq
L^{(O)}_m|\Psi_{phys}\rangle = 0\hskip1,3cm m>0\hskip0,3cm \nonumber\\
L^{(O)}_0|\Psi_{phys}\rangle = (1/2 - a) |\Psi_{phys}\rangle\nonumber\\
G^{(O)}_r|\Psi_{phys}\rangle = 0\hskip1,3cm r\geq 0 \hskip0,4cm
\qqq
In the target space, BCFT's are associated to the scattering of open string whose 
endpoints must obey some boundary condition consistent with the no-flow of energy 
requirement. The boundaries are then mapped to a defect in spacetime. 
Open string perturbative analysis 
will give the formulation of D-branes, a local perturbation on the background geometry.


\subsection{The Cardy program}

\hskip0,5cm It is possible to relate the open string boundary conditions to closed strings by 
considering a one-loop diagram obtained by the periodic identification of the 
worldsheet time $\tau\sim \tau+2\pi T$. By worldsheet duality, we may swap the roles 
of time and space and the diagram becomes a closed string tree-level diagram with 
$2\pi$-periodic space coordinate and boundaries at $t_\alpha=0$ and $t_\beta=\pi/T$. 
This is the first step of the Cardy program, that allows us to relate by worldsheet 
duality the partition function on the annulus with $\alpha$- and $\beta$-boundary 
conditions to the closed string propagating on the cylinder from an inital boundary 
state to a final one:
\qq
Z_{\alpha \beta}(q):=Tr_{\CH_{\alpha \beta}} q^{H^{(O)}}=\langle\alpha |(\tilde q^{1/2})^{H^{(C)}}|\beta \rangle
\label{card}
\qqq
where the trace is taken over the open channel space of states with boundary conditions 
$\alpha$ and $\beta$, associated to the boundary states $|\alpha\rangle$ and $|\beta\rangle$
in the closed channel space of states. The hamiltonians are given by the zero-mode 
Virasoro operators
\qq
H^{(O)}=L_0^{(O)}-c/24\hskip1.0cm H^{(C)}=L_0^{(C)}+\tilde L_0^{(C)}-c/12
\qqq
of the open channel and closed channel respectively.
Finally, the parameters $q=e^{2\pi i T}$ and $\tilde q=e^{-2\pi i/T}$ are related by euclidian worldsheet duality $iT\rightarrow (iT)^{-1}$. 

The closed channel space of states decomposes into a sum of products of left- and 
rigth-chiral sectors
\qq
\CH=\mathop{\oplus}\limits_{\alpha}{\cal H_{\alpha}}\otimes\tilde
{\cal H}_{\alpha}
\qqq
where each chiral space $\CH_\alpha$ carries an irreducible representation of the
chiral algebra: the left-moving (super-)Virasoro algebra or its extension and similarly
for the right-movers (we restrict ourselves to diagonal CFT's).  
There is a single vacuum sector ${\cal H_0}\otimes\tilde{\cal H_0}$ with the vacuum 
state 

\qq
|\Psi_{vacuum}\rangle := |0\rangle|\tilde0\rangle\,.
\qqq
The operator-state correspondence associates to a primary state $\alpha$ in each chiral 
sector a primary field $\phi_\alpha$. The fields obey the fusion algebra
\qq
\phi_\alpha \star \phi_\beta = \sum_i N_{\alpha \beta}^\gamma \phi_\gamma
\qqq
where $N_{\alpha \beta}^\gamma$ are the entries of the fusion matrices that can also be seen as 
the number of copies of the representation labeled by $\gamma$ 
occurring in the open string spectrum
\qq
Z_{\alpha \beta}=\sum_i N_{\alpha \beta}^\gamma\chi_\gamma (q)
\qqq
with $\chi_\alpha(q)$ the character of the representation of the chiral algebra
in $\CH_\alpha$
\qq
\chi_\alpha(q) = {\rm Tr}_{\CH_\alpha} q ^{L^{(O)}_0 - \frac{c}{24}}
\qqq
Under worldsheet duality $q \mapsto \tilde q$, they transform under the modular $S$-matrix
\qq
\chi_\alpha(\tilde q) = \sum_\beta S_{\alpha\beta}\chi_\beta(q)
\qqq
with
\qq
S S^\ast=1\,,\hskip1.0cm S=S^t\,,\hskip1.0cm S^2=C
\qqq
where $C$ is the charge conjugation matrix.
The second step of Cardy was to express the boundary states as linear combinations 
of the so called Ishibashi states
\qq
|\alpha\rangle = \sum_\beta\frac{S_{\alpha \beta}}{\sqrt {S_{0\beta}}}|D_\beta\rangle
\qqq
in the closed channel space of states. The Ishibashi states were defined in \cite{ishibashi} has the unique solution in $\CH$ (up to a normalization factor) of the equation
\qq
(J_n^a + \tilde{J}_{-n}^a)|D_\beta\rangle=0
\qqq
which impose that $|D_\beta\rangle$ does not break the current algebra symmetry. The solution is given by the (formal) expression 
\qq
|D_\beta\rangle=\sum\limits_{\{|e^\beta_n\rangle\}}|e^\beta_n\rangle\otimes|\tilde e^\beta_n\rangle
\qqq
where $\{|e^\beta_n\rangle\}$ is a complete orthonormal basis of $\CH_\beta$. They obey the condition 
\qq
(L_m^{(C)}-\tilde L_{-m}^{(C)})|D_\beta\rangle = 0\hskip1.0cm \forall m
\qqq

The relation (\ref{card}) is proven by the use of Verlinde formula

\qq
N_{\alpha\beta}^\gamma = \sum_l\frac{S_\alpha^\delta S_\beta^\delta 
\bar{S}_\gamma^\delta}{S_0^\delta}
\qqq
 
It is also interesting to write the behaviour of the bulk fields $\phi_\alpha$ 
of conformal weight $h$ when we approach the boundary $z=\bar z$ corresponding to
the boundary state $\alpha$. It is given by the bulk-boundary OPE 
\qq
\phi_\alpha(z)\phi_\alpha(\bar z)=\sum_i(z-\bar z)^{h_i-2h}\phi^{\alpha A\alpha}_i(x)+...
\qqq
where $x=(z+\bar z)/2$ and $\phi^{\beta A\gamma}_i (x)$ denotes boundary  
operators of conformal weight $h_i$ changing the boundary condition $\beta$ 
to $\gamma$ (label A takes care of the multiplicity).


\section{Boundary state formalism}

\hskip0,5cm The open string theory is described by the same sigma model as in the closed string case 
but now with the integration restricted to a worldsheet with boundary that, in the 
Lorentz picture, is parametrized by the coordinates $(\tau, \sigma)$ with metric 
$\gamma_{ab}$ of signature $(-,+)$. The simplest open worldsheet is the timelike
strip with $0\leq\sigma\leq\pi$. The conformal boundary conditions along the timelike
lines $\sigma=0,\pi$ are a combination 
of Neumann and Dirichlet conditions that in the target space are interpreted as free 
open strings and strings stretched between D-branes, respectively. 

If we want to explore time-dependent string theory within the 
CFT on the worldsheet, we may try to explore other types of boundaries - spacelike 
or null - where for each of them we impose Neumann or Dirichlet conditions on the target 
fields. This simple consideration comes from the fact that in special relativity, we
may treat the time and space coordinates similarly, so that all spacetime 
boundaries are at equal footing in our considerations.

So from now on we will use two labels for the lorentzian BCFT to specify the type
of the boundary conditions. The Neumann condition with respect to a timelike boundary 
will be called an $(N;t)$-boundary condition, the Dirichlet condition with respect 
to a null boundary will be called $(D;n)$ and so on.


\subsection{$D$-branes and the non-chiral coherent Ishibashi states}
\hskip0,5cm In this section we describe the boundary states in the closed string sector corresponding to
open strings whose endpoints obey a Neumann or a Dirichlet condition. They will be linear 
combinations of Ishibashi states that fulfil the boundary conditions that we impose 
on the open string endpoints.  We first restrict ourselves to the case of the bosonic string described 
by the following worldsheet action with timelike boundaries 
\qq
S_{open}=\frac{1}{4\pi\alpha'}\int_0^\pi d\sigma\int_{-\infty}^\infty d\tau\,
\partial_a X^\mu\partial^a X_\mu
\qqq
where we have considered the diagonal worldsheet metric $ds^2=-d\tau^2+d\sigma^2$ \cite{pl}.
At timelike boundaries we can either have a Neumann boundary condition $(N;t)$   
\qq
\partial_\sigma X=0\hskip1.0cm (\sigma=0,\,\pi)
\qqq
or a Dirichlet boundary condition $(D;t)$
\qq
\delta X =0 \Leftrightarrow \partial_\tau X=0\hskip1.0cm(\sigma=0,\,\pi)
\qqq 
as these make the boundary contributions to the equation $\delta S_{open}=0$ vanish.
If in the action we restrict the time integral to the interval $[0,T]$ and the field
configurations to the periodic ones $X(\sigma,0)=X(\sigma,T)$ then we may apply  
the worldsheet duality, where space and time are interchanged. Under such 
interchange the Neumann boundary condition and simple rescaling, $(N;t)$ is replaced 
by the $(N;s)$-condition 
\qq
\partial_\tau X=0\,\,\,\,at\,\,\tau=0,\pi/T
\qqq
and the Dirichlet boundary condition $(D;t)$ by the $(D;s)$-condition
\qq
\partial_\sigma X=0\,\,\,\,at\,\,\tau=0,\pi/T
\qqq
in the closed string sector. 

The solutions of the above equations are nicely related by noting that 
the introduction of boundaries allow us to relate the $U(1)$-chiral 
currents 
$J(x^+)=\partial_+X_L(x^+)$ and $J(x^-)=\partial_-X_R(x^-)$  
of the bulk theory by a gluing map $\Omega$ 
\qq
J(x^-)=\Omega J(x^+)
\qqq
on the boundary.
Considering the Fourier expansion of the chiral bosonic fields
\qq
X_L(x^-)={\hat x\over 2}+\alpha_0x^-\,+\,{1\over 2}\sum_{n \neq 0}
{\alpha_n \over n}e^{-inx^-}\nonumber\\
X_R(x^+)={\hat x\over 2}+\tilde\alpha_0x^+\,+\,{1\over 2}
\sum_{n \neq 0}{\tilde{\alpha}_n \over n}e^{-inx^+}\,.
\qqq
and substituting them in the definition of the $U(1)$ currents,
we find that $\Omega=1$ for Neumann condition and $\Omega=-1$ for the Dirichlet boundary 
condition at $\sigma=0,\pi$ and the other way around for $\tau=0,\pi/T$,
with the coupling of the chiral modes 
\qq
\alpha_n + \Omega \tilde{\alpha}_{-n}\,=\,0\,\,\,\,\forall n\neq 0
\qqq
and the field solution  $X(\tau,\sigma)=X(x^+)+ X(x^-)$. 
These conditions were used in \cite{callan} to define the boundary state associated with them.

Consider the example of a $D_p$-brane defined by imposing Neumann conditions on its 
$p$ longitudinal coordinates and Dirichlet conditions at the $D-p$ transverse coordinates. 
We define the boundary state as a sum of Ishibashi states in the closed string 
sector satisfying
\qq
\partial_\tau X^\alpha \mid_{\tau=0}|B\rangle_X\,=\,0\,\,\,\,\,\,\,\alpha=0,...,p\nonumber\\
X^i\mid_{\tau=0}|B\rangle_X\,=\,y^i\,\,\,\,\,\,\,i\,=\,p+1,...,D-1
\qqq
that in terms of the closed string oscillators is translated on
\qq
(\alpha_n^\mu\,+\,S^\mu_\nu\tilde\alpha_{-n}^\nu)|B\rangle_X\,=\,0\,\,\,\,\forall n\neq 0
\label{bounc0}\nonumber\\
\hat p^\alpha|B\rangle_X\,=\,0\hskip1,5cm\label{bounc1}\nonumber\\
(q^i-y^i)|B\rangle_X\,=\,0\hskip1,5cm
\label{bounc}
\qqq
where we have introduced the matrix
\qq
S^{\mu\nu}=(\eta^{\alpha\beta},-\delta^{ij})\hskip0,3cm .
\qqq
The eigenstates of (\ref{bounc0}) are obtained, up to a normalization 
factor, by a Bogoliubov transformation 
\qq
|B\rangle_X\,=\,N_p\,\delta^{25-p} (q^i-y^i)\,\exp \left(-\sum_{n=1}^{\infty}\frac{1}{n}\alpha_{-n} \cdot S \cdot \tilde \alpha_{-n}\right)|0\rangle|\tilde 0\rangle|p=0\rangle
\qqq
To those states one can add the Majorana field contributions as well as the bosonic and fermionic Faddev-Popov ghosts
\qq
|B\rangle_\psi\,=\,\exp\left(\pm i\sum_{r>0} \psi_{-r}^\mu\tilde \psi_{-r}^\mu\right)|0\rangle\hskip5,0cm\nonumber\\
|B\rangle_{gh}\,=\,\exp \,\huge(\,\sum_{m=1}^\infty [c_{-m}\tilde b_{-m}\,+\,\tilde c_{-m}b_{-m}]\pm\hskip3,5cm\nonumber\\ 
\hskip3,5cm i\sum_{r>0}[\gamma_{-r}\tilde\beta_{-r}-\tilde\gamma_{-r}\beta_{-r}]\,\huge)\,\frac{1}{2}(c_0+\tilde c_0)|\downarrow\downarrow\rangle
\qqq
where $|\downarrow\downarrow\rangle$ is annihilated by all positive frequency ghost and superghost oscillators and the anti-ghost zero-modes.
The above states are the Bogoliubov solutions of the boundary conditions
\qq
\psi_r^\mu=\pm i\tilde\psi_{-r}^\mu\hskip3,4cm\nonumber\\
c_m=-\tilde c_{-m}\hskip1,0cm
b_m = \tilde b_{-m}\hskip0,8cm\nonumber\\
\gamma_r=\mp i\tilde\gamma_{-r}\hskip1.0cm
\beta_r=\mp i\tilde \beta_{-r}\hskip0,3cm .
\qqq

The full boundary state 
$|B\rangle\,=\,|B\rangle_X\,|B\rangle_\psi\,|B\rangle_{gh}$
is a sum of {\bf coherent} and {\bf nonchiral} Ishibashi states.
The imposition of no flux-energy crossing non-null boundaries is translated 
to the condition of conformal invariance
\qq
(L_n^{(C)} - \tilde L^{(C)}_{-n})|B\rangle=0\hskip1.0cm 
\qqq
that comes from $\alpha_n+\Omega\tilde\alpha_{-n}$ with $\Omega^2=1$.
This was interpreted in \cite{callan} as a diffeomorphism invariance condition under the 
group $Diff(S^1)$ of diffeomorphisms of the circle. Explicity, we define the operator
$S_n=L_n - \tilde L_{-n}$ that can be seen to obey the algebra
\qq
[S_n, S_m] = (n-m) S_{n+m}
\qqq
coming from the Virasoro algebra after we impose $c=\tilde c$. 
\vskip0.5cm
{\bf Summarizing:} Boundary states must obey two conditions, 
a no-flux of energy crossing the physical boundary and a conformal $Diff(S^1)$ 
invariance that in the present case coincide.


\subsection{Deformed Ishibashi states and the one-dimensional path integral approach}

One would like to calculate the boundary states for a $D_p$-brane immersed in 
a background gauge field $A_\mu(X)$ with an interaction given in the closed 
string channel at $\tau=0$ by
\qq
S_A=\frac{1}{4\pi}\int_0^{2\pi}d\sigma \left(A_\mu(X)\partial_\sigma X^\mu - \frac{i}{2}F_{\mu \nu}(X)\theta^\mu\theta^\nu \right)
\qqq
in the euclidian picture, where $F_{\mu\nu}=\partial_\mu A_\nu - \partial_\nu A_\mu$ and $\theta^\mu=\psi^\mu\pm i\tilde\psi^\mu$ for R sector or NS sector.
The bosonic boundary conditions in the closed string sector become a mixing of Neumann and Dirichlet conditions 
\qq
\partial_\tau X_\alpha\, + \,F_{\alpha\beta}\partial_\sigma X^\beta = 0
\qqq
at $\tau=0,\pi/T$. How does this affect the previous boundary states? 

To answer this question 
we consider the example of a simple harmonic oscillator. Let us search for the oscillator 
state satisfying for a given value of the momentum $p$ the eigenvalue condition 
\qq
(a - a^\dag  - ip)|p\rangle=0
\label{mlc}
\qqq
where $a$ and $a^\dag$ are respectively the
annihilation and creation operators.
We want as well that the solution  obey the completeness relation
\qq
\int_{-\infty}^\infty dp |p\rangle\langle p| = 1
\qqq
After some algebra we find the eigenstates
\qq
|p\rangle = (2\pi)^{-1/4}e^{-p^2/4}e^{+(a^\dag)^2/2 \,+\, ipa^\dag}|0\rangle
\qqq
Note that after integration over all posible momentum values, we find a non-normalized squeezed state 
\qq
|\Psi\rangle = \int  dp\,|p\rangle = (8\pi)^{1/4} e^{-(a^\dag)^2/2}|0\rangle
\qqq
as we have chosen from the beginning a state perfectly localized in momentum space, see (\ref{mlc}). On the other hand, the insertion of a perturbative action $S_{pert}(p)$ at a given time serves to create the perturbed oscillator state 
$|\Psi_{pert}\rangle$ with 
\qq
|\Psi_{pert}\rangle = \int dp\,e^{-S_{pert}(p)}|p\rangle
\qqq 
In the string case, the boundary perturbation at a given time will change the free boundary states by a multiplication of a Wilson line factor
\qq
{\rm Tr} \,P \,\exp(-S_A)
\qqq
in the Polyakov path integral where $P$ denotes path (integral) ordering operator.
We thus obtain the perturbed state as
\qq
|A\rangle\,=\, \int  \CD\bar p \CD p \CD\bar\Theta \CD\Theta\,{\rm Tr}\,P\,\exp(-S_A)|p,\bar p\rangle|\Theta, \bar \Theta;\pm\rangle
\label{ptt}
\qqq
where $|p,\bar p\rangle$ are bosonic eigenstates associated to the free theory of two commuting set of bosonic coordinates as we will soon see. The same holds for the Majorana contribution $|\Theta, \bar \Theta;\pm\rangle$ associated to two anticommuting sets of fermionic fields.

To calculate the bosonic eigenstates we consider the closed string as a combination of left- and right-oscillators created at $\tau=0$
\qq
X^\mu(\sigma,0)= q^\mu+(\alpha')^{1/2}\sum_{m\neq 0} |m|^{-1/2}\left[a_m^\mu e^{-im\sigma}+\tilde a_m^\mu e^{im\sigma}\right]
\label{fX2}
\qqq
for which we impose the Dirichlet condition 
\qq
\partial_\sigma X^\mu=0\hskip1.0cm (\tau=0)
\qqq
so that the left/right oscillator coupling is given by $a_m^\mu\,+\,\tilde a_{-m}^\mu\,=\,0$, for all $m\neq 0$.
In this way, the target coordinate with a Dirichlet boundary condition is given by combinations of the following two sets
\qq
ip_m^\mu = a_m^\mu - \tilde a_m^{\mu \dag}\nonumber\\
-i\bar p_m^\mu = a_m^{\mu \dag} - \tilde a_m^\mu
\qqq
for $m>0$ where we have denoted $a^\dag_m=a_{-m}$. These two combinations form a complete commuting set of bosonic oscillators.  As in the one-dimensional oscillator case, we interpret these equations as eigenvalue conditions that should define certain eigenstates $|p,\bar p\rangle$. After we impose the completeness condition on them
\qq
\int \CD p \CD\bar p |p,\bar p\rangle\langle p , \bar p|\,=\,1
\qqq
we find 
the solution  
\qq
|p,\bar p\rangle = \exp\left[-\frac{1}{2}(\bar p|p)+(a^\dagger|\tilde a^\dagger)+i(a^\dagger|p)+i(\bar p|\tilde a^\dagger)\right]|0\rangle|\tilde 0\rangle
\qqq
where we  use the notation
\qq
(\bar p|p)=\sum_{\mu=0}^{D-1}\sum_{m=1}^\infty \bar p_m^\mu p_m^\mu
\qqq
with $D=26$ for the bosonic case or $D=10$ for the superstring case. 
Similar considerations hold if we had considered Neumann boundary conditions \cite{callan}.

The same analysis holds for the Majorana contribution by writing the set of two  anticommuting eigenstates
\qq
(\bar\Theta_r^\mu-\psi_r^{\mu \dag}\mp i\tilde\psi_r^\mu)|\Theta,\bar \Theta;\pm\rangle = 0\nonumber\\
(\Theta_r^\mu-\psi_r^\mu \pm i\tilde\psi_r^{\mu\dag})|\Theta,\bar \Theta;\pm\rangle = 0
\qqq
for $r>0$ and with solution
\qq
|\Theta,\bar \Theta;\pm\rangle = \exp\left[-\frac{1}{2}(\bar \Theta|\Theta)\pm i(\psi^\dag|\tilde\psi^\dag)+(\psi^\dag|\Theta)\mp i(\bar\Theta|\tilde\psi^\dag)\right]|0;\pm\rangle
\qqq
that obeys the completeness relation.

Insert these states corresponding to Dirichlet and Neumann boundary conditions in the path integral (\ref{ptt}) and perturb the system by a constant external field. In this case, the perturbed boundary term reduces to a quadratic form
\qq
S_{A}=\frac{i}{8\pi}F_{\mu\nu}\int_0^{2\pi}d\sigma\left(X^\mu\partial_\sigma X^\nu - i\theta^\mu\theta^\nu\right)
\qqq
so that the functional integrals in (\ref{ptt}) become gaussian. In what follows, we are interested on the case of vanishing gauge fields in the directions transverse to the $D_p$ - brane. After a gaussian integration, the bosonic and fermionic contributions to the perturbed boundary state $|A\rangle$ are \cite{callan}
\qq
|B_X\rangle = \sqrt{-{\rm det}(\eta + F)}\,\delta^{D-p-1}(q - y)\,\exp\left(-\sum_{n=1}^\infty\frac{1}{n}\alpha_{-n}^\mu M_{\mu \nu} \tilde\alpha_{-n}^\nu\right)\,|0\rangle\nonumber\\
|B_\psi\rangle=|B_{R}\rangle = \frac{i}{\sqrt{-{\rm det}(\eta+F)}}\, \exp\left(i\sum_{n=1}^\infty \psi_{-n}^\mu M_{\mu\nu}\tilde\psi_{-n}^\nu\right)\,|0_R\rangle\hskip1,4cm\nonumber\\
|B_\psi\rangle=|B_{NS}\rangle = -i\,\exp\left(i\sum_{n=1/2}^\infty \psi_{-n}^\mu M_{\mu \nu}\tilde\psi_{-n}^\nu\right)|0_{NS}\rangle\hskip2,9cm
\qqq
where 
\qq
M_{\mu\nu}=\left(\left[\frac{1-F}{1+F}\right]_{\alpha\beta},-\delta_{ij}\right)
\label{mM}
\qqq
and we have replaced the $a_n$ oscillator modes by the $\alpha_n$, compare (\ref{fX2}) with (\ref{fX}). 

The above states satisfy the boundary conditions for the bosonic contribution
\qq
\hat p^\beta|B_X\rangle = 0\hskip0.5cm ,\,\,\, q^i|B_X\rangle = y^i|B_X\rangle\hskip0.5cm ,\,\,\,(\alpha_n^i-\tilde \alpha_{-n}^i)|B_X\rangle=0\nonumber\\
\left[(1+F)_{\alpha\beta}\,\alpha_n^\beta \,+\, (1-F)_{\alpha\beta}\,\tilde\alpha_{-n}^\beta\right]|B_X\rangle = 0
\qqq
and the fermionic contribution
\qq
(\psi_n^i\,\pm\,i\tilde\psi_{-n}^i)|B_\psi\rangle = 0\nonumber\\
\left[(1+F)_{\alpha\beta}\,\psi_n^\beta \,\mp\, i(1-F)_{\alpha\beta}\,\tilde\psi_{-n}^\beta\right]|B_\psi\rangle = 0
\qqq
Here we have labelled the indices of transverse coordinates to the $D_p$ - brane by Latin letters and the longitudinal coordinates by Greek letters.

It is interesting to note in (\ref{mM}) that there is a relative boost of left and right-movers as the external gauge field approaches to a critical limit $F\rightarrow 1$. Here there is a complete decoupling of left- and right-chiral sector, as we can see in the above boundary conditions along the transverse directions to the brane
\qq
F\rightarrow 1\hskip1.0cm\Rightarrow\hskip1.0cm \Omega\rightarrow 0
\qqq
Moreover the normalization factors of the above coherent and non-chiral perturbed boundary states become singular at this critical limit.
 We will later show that the alternative boundary states associated to the decoupling limit are {\bf chiral} and {\bf squeezed} states, solutions of Dirichlet conditions with respect to worldsheet null boundaries.            


\section{Calculation of the $D$-brane tension}

\hskip0,5cm The normalization factor $N_p$ in the front of the boundary state associated to the $D_p$-brane is still unknown. To calculate it we consider a Cardy type program in the target space of string theory. 
 For the sake of simplicity we restrict ourselves to the case of bosonic strings and starting to neglect the ghosts contribution. First compute the one-loop free energy of an open string streching bettwen the two $D_p$-branes, given by the Coleman-Weinberg formula \cite{PolCJ}
\qq
F=\int_0^\infty \frac{dT}{2 T}{\rm Tr}_{n,k}\left[ e^{-2\pi T(L_0^{(O)}-1)}\right]\nonumber\\
\,\,=\int_0^\infty\frac{dT}{2T}{\rm Tr}_{n,k}\left[ e^{-2\pi T\alpha'(k^2+M^2)}\right]
\label{cwf}
\qqq
where $T$ parametrizes the periodic time variable $T\sim T+2\pi$ and $L_0^{(O)}$ is the Virasoro operator on the open string sector related to the mass spectrum 
\qq
M^2=\frac{1}{\alpha'}\left(\sum_{n=1}^\infty \alpha_{-n}\cdot\alpha_n\,-\,1\right)+\frac{|y|^2}{4\pi^2\alpha'^2}
\label{massf}
\qqq
and $|y|$ is the value of the distances between the two $D_p$-branes.
The trace ${\rm Tr}_{n,k}$ is to be read as a integration over the momentum along the brane and a trace over the oscillator modes. After substitution of (\ref{massf}) into (\ref{cwf}) we have
\qq
F=2V_{p+1}\int_0^\infty\frac{dT}{2T}(8\pi^2\alpha'T)^{-\frac{p+1}{2}} e^{-\frac{|y|^2 T}{2\pi\alpha'}}f^{-26}_1(e^{-\pi T})
\qqq
with the front factor of two coming from the freedom of changing the oriented string endpoints, and
\qq
f_1^{-1}(e^{-\pi T}):=\prod_{n=1}^\infty\left(\frac{1}{1-e^{-2\pi T n}}\right)={\rm Tr}_n\prod_{n=1}^\infty e^{-2\pi T\alpha_{-n}^\mu\alpha_n^\mu}
\qqq
without summation on the index $\mu$. Under the worldsheet modular transformation $T\rightarrow \tau=1/T$, the function $f_1$ has the property
\qq
f_1(e^{-\pi T})=\sqrt \tau f_1(e^{-\pi \tau})\hskip0,3cm .
\label{mpr}
\qqq
The result should be compared to
the free-energy $F=\,\langle B_X|D|B_X\rangle$ of a closed string created at one brane and propagating freely until anihilated by the second brane. Under the change of variable $z=e^{-i\pi \tau}$, the cylinder amplitude may be calculated using the disk operator
\qq
D_a=\frac{\alpha'}{4\pi}\int_{|z|\leq 1} \frac{d^2 z}{|z|^2}z^{L_0^c-a}\bar z^{\tilde L_0^c-a}\hskip0,3cm ,
\qqq
which is the closed string propagator written in terms of the Hamiltonian given by the closed string zero-mode Virasoro operators $H^{(C)}=L_0^{(C)}+\tilde L_0^{(C)}$. The constant $a$ comes from normal ordering and in the case at hand we set $a=1$. We reproduce here the bosonic boundary state for the bosonic string
\qq
|B_X\rangle=N_p\,\delta^{25-p}(q^i-y^i)\prod_{n=1}^\infty e^{-\frac{1}{n}\alpha_{-n}^\mu S_{\mu \nu}\tilde\alpha_{-n}^\nu}|0\rangle|\tilde 0\rangle|p=0\rangle
\qqq
and the Virasoro operators
\qq
L_0^{(C)}=\frac{\alpha'}{4}\hat p^2+\sum_{n=1}^\infty\alpha_{-n}\cdot\alpha_{n}\hskip0.5cm ,\,\,\,\tilde L_0^{(C)}=\frac{\alpha'}{4}\hat p^2+\sum_{n=1}^\infty\tilde\alpha_{-n}\cdot\tilde\alpha_{n}\hskip0,3cm .
\qqq
By substitution in the amplitude, we first split the zero mode from the non-zero oscilator mode terms. The non-zero modes give the contribution
\qq
f^{-26}_1(|z|)=\prod_{n=1}^\infty\left(\frac{1}{1-|z|^{2n}}\right)^{26}\hskip0,3cm .
\qqq
The zero mode contribution will be the important one to obtain the value of the normalization factor, so we will be examine it carefully 
\qq
A_0 = \langle p=0|\delta^{25-p}( q^i)|z|^{\frac{\alpha'}{2}\hat p^2}\delta^{25-p}( q^i-y^i)|p=0\rangle\hskip0,3cm .
\qqq
Writing $A_0$ as a Fourier transformation we get
\qq
A_0 = \int\int\frac{d^{25-p}k}{(2\pi)^{25-p}}\frac{d^{25-p}k'}{(2\pi)^{25-p}}\langle p=0|e^{ik\cdot\hat q}|z|^{\frac{\alpha'}{2}\hat p^2}e^{ik'\cdot(q - y)}|p=0\rangle
\qqq
where $k$ is the one-loop transverse momentum.
Using the identities
\qq
e^{ik\cdot\hat q}|p=0\rangle\,=\,e^{ik\cdot\hat q}|p_\perp=0\rangle|p_\parallel=0\rangle\,=\,|p_\perp=k\rangle|p_\parallel=0\rangle\nonumber\\
\hat p^2|p_\perp=k\rangle|p_\parallel=0\rangle\,=\,k^2|p_\perp=k\rangle|p_\parallel=0\rangle\nonumber\\
\langle p=k||p=k'\rangle\,=\,2\pi\delta(k-k')\nonumber\\
V_d:=(2\pi)^d\delta^d(0)
\qqq
and changing the variable  $\tau$, we end up with the gaussian integral
\qq
A_0 = V_{p+1}\int\frac{d^{25-p}k}{(2\pi)^{25-p}}e^{-\frac{\pi}{2}\tau \alpha' k^2+ik\cdot y}\nonumber\\ \\
=V_{p+1}e^{-y^2/(2\pi\tau\alpha')}(2\pi^2\tau\alpha')^{(25-p)/2}\nonumber
\qqq
The total contribution 
\qq
\langle B_X|D_1|B_X\rangle =  (N_p)^2 V_{p+1}\frac{\pi\alpha'}{2}\int_0^\infty d\tau (2\pi^2\tau\alpha' )^{-(25-p)/2}e^{-y^2/(2\pi\tau\alpha')}f^{-26}_1(e^{-\pi \tau})\nonumber
\qqq
should be compared to the one-loop open string calculation by performing the worldsheet duality 
$\tau=1/T$ and using the modular property of $f_1$, see (\ref{mpr}). The ghost contribution corrects 
the power by replacing $26$ with $24$. After this remark, we are able to find 
the normalization factor in the boundary state of the $D_p$ brane 
\qq
N_p=\frac{\sqrt \pi}{2} T_p\hskip0.3cm ,\hskip0,6cm T_p=\frac{(2\pi\sqrt{\alpha'})^{11-p}}{2^4}\hskip0,3cm .
\qqq
As shown in \cite{DiVecchia2,DiVecchia3} we can read from the boundary state the meaning 
of certain  coupling constants on the low-energy effective field theory by taking projections
\qq
C_\Psi \sim \langle \Psi|B_X\rangle
\qqq
where $|\Psi\rangle$ denotes closed string state associated to the massless field in question. In particular 
\qq
A^{\mu \nu}:=\langle0;k|\alpha_1^\mu\tilde\alpha_1^\nu|B_X\rangle=-\frac{T_p}{2}V_{p+1}S^{\mu\nu}
\qqq 
and by saturating it with the symmetric graviton entering in the vertex operator (\ref{gvo})
\qq
A_{grav}=A^{\mu\nu}\epsilon_{\mu\nu}=-T_p V_{p+1}\eta^{\alpha\beta}\chi_{\alpha\beta}\hskip0,3cm .
\qqq
This is interpreted as the value of the tension (energy per unit brane volume) 
due to exchange of gravitons with the brane.


\section{The Dissipative Hofstadter Model and Open Strings}

\hskip0,5cm It is known that open strings immersed in a constant electromagnetic background may be 
described by one dimensional field theory of dissipative quantum mechanics at the 
critical point. A particular case is the dissipative Hofstadter model that we shall 
treat here.
Consider a particle trajectory described by the position vector $\vec X$ immersed in 
a thermal bath of an infinite number of harmonic oscillators with frequency 
$\omega_\alpha$ coupled linearly to $\vec X$ with a given coupling constant 
$C_\alpha$. This can be described macroscopically as a dissipative force of the 
form $-\eta \vec{\dot X}$ where $\eta$ is the coefficient of friction. Such 
macroscopic 
friction may be rendered microscopically by the relation
\qq
\eta\omega=\sum_\alpha\frac{C_\alpha^2}{\omega_\alpha}\delta(\omega-\omega_\alpha)
\qqq
where $\omega$ is the frequency of $\vec X$. This is known as the Caldeira-Leggett model, 
a particular Dissipative Quantum Mechanics model (DQM for short). The DQM models convert 
quantum mechanics into a one-dimensional statistical mechanics problem. Such models  
becomes interesting when applied to the quantum mechanics of an electron moving in 
two dimensions subject to a uniform magnetic field and a periodic potential, 
usualy called the Hofstadter model. The introduction of the Caldeira-Leggett thermal 
bath to the  Hofstadter model will define the Dissipative Hofstadter Model (DHM). 

The Lagrangian of such system is given by 
\qq
S_{DHM}=\int_0^T dt\left[\frac{1}{2}M\dot{\vec X}^2+V(\vec X)+iA_j(\vec X)\dot X^j\right]\,+\hskip1,0cm\nonumber\\  \frac{\eta}{4\pi}\int_0^T dt\int_{-\infty}^\infty dt'\frac{(\vec X(t)-\vec X(t'))^2}{(t-t')^2}\nonumber
\qqq
where $V$ is a scalar potential and $A$ is the vector potential of the magnetic field. 

In the euclidian path integral 
\qq
Z(\eta, B, V, 0)=\int D X(t)\,e^{-S_{DHM}}
\qqq
we may add a linear source term 
\qq
S_F=\int \vec F(t)\cdot\vec X(t)\,dt
\qqq
and define the generating functional 
\qq
Z(\eta, B, V, F)=\int D X(t)\,e^{-S_{DHM}-S_F}
\qqq
that allows to compute the correlation functions. In particular, the two-point 
correlation function is 
\qq
\langle X^\mu(t_1)X^\nu(t_2)\rangle = \frac{1}{Z(\eta,B,V,0)}\frac{\delta^2Z(\eta,B,V,F)}{\delta F_\mu(t_1)\delta F_\nu(t_2)}_{\line(0,1){25}{F=0}}\hskip0,3cm .
\qqq
The kinematic term is quadratic and therefore irrelevant in the renormalization group 
so we set $M=0$. Consider the case of the charged particle immersed in a 
magnetic field given by the linear gauge
\qq
(A_x,\,A_y)\,=\,\frac{1}{2}(By,\,-Bx)
\qqq 
and take the periodic potential 
\qq
V(x,y)\,=\,V_0\cos(2\pi x/a)\,+\,V_0\cos(2\pi y/a)\hskip0,3cm .
\qqq
In this case the two-point correlation function is 
\qq
\langle X_i(t)X_j(t')\rangle=-\frac{\alpha}{\alpha^2+\beta^2}\log(t-t')^2\delta_{ij}-i\frac{\pi\beta}{\alpha^2+\beta^2}{\rm sign}(t-t')\epsilon_{ij}
\qqq
for $\alpha=\eta a^2/2\pi\hbar$ and $\beta=eBa^2/2\pi\hbar c$ where $\hbar$ is the Planck constant and $c$ is the speed of light.

The first term on the right hand side measures the delocalization of the particle
wave function, called mobility in the condensed matter literature. 
The logarithmic growth is a 
transition between two extreme limits defined by the long-time behaviour: bound 
by a constant or growing without limit. The coeficient at the front of the 
logarithm is the value of the critical mobility. 

The second term can be interpreted as a Hall effect, it measures the response 
of the system in the transverse directions to the applied magnetic field.


\section{Boundary Deformation Theory}

Given a closed string background described by a bulk CFT one may ask what are 
the possible D-brane configurations associated to boundary CFT's. It turns out that one 
has multiple choices parametrized by marginal boundary fields, forming the D-brane 
moduli space for a given bulk background.
There may be more symmetries on BCFT than those generated by the Virasoro algebra. 
The most general symmetries are generated by certain currents forming a $\CW$ 
algebra, where the  Virasoro is a subalgebra of $\CW$.  By the usual method of 
images we may construct the general chiral algebra in the open channel by computing 
the chiral currents
\qq
W_n^{(O)}:=\frac{1}{2\pi i}\int z^{n+h_W-1}W(z)dz\,-\,\frac{1}{2\pi i}\int \bar z^{n+h_W-1}\Omega(\tilde W)(\bar z)d\bar z
\qqq 
where the integrals are performed along a semi-circle in the upper half plane. 
Here $h_W$ is the conformal height of the operator $W$ and $\Omega$ denotes the automorphism relating left-movers to the right-movers.

Marginal boundary fields are introduced in the free system as boundary perturbations that will change the boundary conditions given by the gluing map $\Omega$ of the current algebra, but without affecting the local bulk properties of the system.
An example of that is the Dissipative Quantum Mechanics that we have described in the previous section.

Depending on the value of  the conformal weight of the boundary field, we may distinguish
three general behaviours of the renormalization group flow under the boundary perturbation. 
For $h>1$ the perturbative field is said to be {\it irrelevant} as the RG-flow leads 
to the same original BCFT. In the {\it marginal} case ($h=1$) we do not introduce 
any length scale in the theory so that under RG-flow we most probably end up in the same 
fixed point of the theory or in another point of the fixed point manifold.  
The most interesting case arises when $h<1$ where the perturbations are {\it relevant}. 
In such case, it is difficult to follow the RG-flow, which will end up in some unknown 
fixed point of the theory. We stress here that it is assumed that the perturbations 
do not affect the local bulk properties. 

A more systematic treatment was done in \cite{RecSch} where the authors consider two 
types of marginal deformations - chiral and non-chiral. In chiral deformations branes 
(BCFT) are related to each other by continuous symmetries on the target space. 
In the second case, non-chiral deformations can for example push the brane to 
some singularity on the target space. 

The theory is perturbed by the boundary operator 
\qq
I_{\lambda \Psi} = P\,\exp\{\lambda S_\Psi\} 
= P\,\exp\{\lambda \int_{-\infty}^\infty \Psi(x)\frac{dx}{2\pi}\}
\qqq
where $P$ denotes path ordering in the Polyakov path integral and we are using 
the complex plane where the boundary is located at the real line. For simplicity 
we impose that the boundary field $\Psi(x)$ should be mutual- (and self-) local 
with respect to another boundary field $\Phi(x')$, that is
\qq
\Psi(x_1)\Phi(x_2)=\Phi(x_2)\Psi(x_1)\hskip1.0cm (x_1\,<\,x_2)\hskip0,3cm .
\qqq

An example of a chiral marginal perturbation is the one given by the  $U(1)$ current $J$  
for which the gluing map is deformed as 
\qq
W(z)\,=\,\Omega \circ \gamma_{\bar J}(\bar W)(\bar z)\hskip1.0cm (z\,=\,\bar z)
\qqq
with 
\qq
\gamma_J(W)\,=\,\exp(-i\lambda J_0)W\exp(i\lambda J_0)
\qqq
an inner automorphism of the chiral $\CW$ algebra. In particular, the inner automorphism 
acts trivial on the Virasoro field leading to the condition $T=\bar T$ interpreted 
as no flow of energy across the timelike boundaries $\sigma=0,\pi$.  This last condition 
continues to be valid even if we deforme the theory by self-local non-chiral marginal 
boundary fields. For the other currents the gluing map is deformed by general 
local marginal boundary fields as 
\qq
W(z)e^{\lambda\int\frac{dx}{2\pi}\psi(x)}\,=\,e^{\lambda\int\frac{dx}{2\pi}\psi(x)}\left[e^{\lambda \psi}\tilde W\right](\bar z)
\qqq
with
\qq
\left[e^{\lambda \psi}\tilde W\right](\bar z):=\sum_{n=0}^\infty \frac{\lambda^n}{n!}\int_{\gamma_1}\frac{dx_1}{2\pi}...\int_{\gamma_n}\frac{dx_n}{2\pi}
\tilde W(\bar z_\delta)\psi(x_1)...\psi(x_n)
\qqq
where we have used $\bar z_\delta = z-2i\delta$ for $\delta$ a positive infinitessimal real parameter and the integrations are taken over straight lines $\gamma_i$ at the upper half plane, getting closer and closer to the real line as $i$ runs from $1$ to $n$. 

An example of non-chiral self-local deformation was developed in \cite{CKLM} 
where to a free field we add a boundary perturbation given by a periodic 
complex potential with strength $|\lambda|$. The system is described by the Lagrangian
\qq
L=-\frac{1}{8\pi}\int d\tau d\sigma \,\left((\partial_\tau X)^2\,-\,(\partial_\sigma X)^2\right) \,-\,\frac{1}{2\sqrt 2 \pi}\int d\tau \left( \lambda e^{iX(0)/\sqrt 2}+\bar \lambda e^{-iX(0)/\sqrt 2}\right)\nonumber
\qqq
where $X$ is the target coordinate that parametrizes an open string. By imposing a Dirichlet boundary condition at $\sigma=\pi$, the boundary condition at the first endpoint $\sigma=0$ becomes dynamical
\qq
-\partial_\sigma X+i\lambda e^{iX/\sqrt 2}-i\bar \lambda e^{-iX/\sqrt 2}=0
\qqq
and by varying $|\lambda|$ from zero to infinity, we are able to interpolate between Neumann to Dirichlet boundary conditions. More systematically, consider the case when $\lambda$ is real. The perturbation is generated by two boundary fields
\qq 
\psi^1(x)=\left(e^{i\sqrt 2 X(x)}+e^{-i\sqrt 2 X(x)}/\sqrt 2\right)\nonumber\\
\psi^2(x)=\left(e^{i\sqrt 2 X(x)}-e^{-i\sqrt 2 X(x)}/\sqrt 2\right)\,\,,
\qqq
that are local with respect to themselves. It can be seen that the gluing condition in the energy momentum tensor is unchanged $T(z)=\tilde T(\bar z)$ but on the currents $(J,\tilde J)=(\partial X,\bar\partial X)$, the above boundary fields will deform the initial gluing map $J(z)=\tilde J(\bar x)$ as follows; under the perturbation by $\psi^1$ we have
\qq
J(z)\,=\,\sin(\sqrt 2 \lambda)\psi^2(x)+\cos(\sqrt 2\lambda)\tilde J(\bar z)
\qqq
and by $\psi^2$ perturbation
\qq
J(z)\,=\,-\sin(\sqrt 2 \lambda)\psi^1(x)+\cos(\sqrt 2\lambda)\tilde J(\bar z)
\qqq
for $z=\bar z = x$. In particular, for $\lambda=(2n+1)\frac{\pi}{\sqrt 2}$, the original Neumann conditions for $J$ turn into Dirichlet conditions.


\chapter{Electrically charged open string and boosted $D$-branes}

\hskip0,5cm It is known that there is pair production of particles in a Rindler spacetime describing 
a constantly accelerated particle detector in a Minkowskian background. 
The particle production comes from the existence of an external force to keep the
motion of the detector, creating a thermal bath around it of a given temperature 
proportional to the acceleration \cite{BD}. There is a critical limit associated 
to the relativistic velocity of the particle.
A second example of pair production is the case of a Minkowskian spacetime with 
an electric field. Nevertheless, there is no critical behaviour as we raise the 
electric field strength.

The existence of an upper limit in the presence of a constant electric field was found 
in the context of string theory. Consider a bosonic open string whose endpoints 
carry an electric charge $e$. The system is described by the following action
\qq
S\,=\,-\frac{1}{4\pi\alpha'}\int d\sigma d\tau \partial_aX^\mu\partial^aX_\mu\,
+\,\frac{e}{2}\int d\tau F_{\mu\nu}X^\nu\partial_\tau X^\mu
\qqq
where the string tension $T$ is related to the Regge slope by $T=(2\pi\alpha')^{-1}$.  
Classically, above the critical electric field $E_{crit}=(2\pi\alpha'e)^{-1}$, the 
tension can not hold the string together \cite{burgess}.
  
A more refined argument was presented in \cite{bachasporrati}, where they make 
a Schwinger's type of calculation on the pair production of open unoriented 
bosonic strings in the presence of a constant electric field ($F_{01}=E$). They 
concluded that the rate of production diverges when the electric field approaches 
to the critical value $E_{crit}$. For future reference we shall describe the main results
on the quantization of the system \cite{fr}. Since we are in the case of a pure 
constant electric field, only the light-cone coordinates involving the direction
of the field are affected. The boundary conditions become a mixture of Dirichlet 
and Neumann conditions
\qq
\partial_\sigma X^\pm\,=\,2\pi\alpha' e E\partial_\tau X^\pm
\qqq
at $\sigma=0,\pi$. In this way, the light-cone coordinates have the mode expansion
\qq
X^\pm(\sigma,\tau)\,=\, \hat x^\pm\,+\,ia_0^\pm\phi_0^\pm\,+\,i(\alpha')^{1/2}\sum_{n=1}^\infty (a_n^\pm\phi_n^\pm(\epsilon)\,-\,h.c)
\qqq
where 
\qq
\phi_n^\pm(\epsilon)\,=\,(n\pm i\epsilon)^{-1/2}e^{-i(n\pm i\epsilon)\tau}
\cos\left[(n\pm i\epsilon)\sigma\mp i\pi\epsilon/2\right]
\qqq
are a complete set of normalized mode functions that satisfy the wave equation and the above boundary conditions. They depend on the parameter $\epsilon=2\,{\rm arctanh}(2\pi\alpha' e E)/\pi\,$. The inner product is defined by the following expression
\qq
\int_0^\pi\frac{d\sigma}{\pi}\bar{\phi_n}^\pm(\tau,\sigma)\left[i\stackrel{\leftrightarrow}{\partial_\tau} + 2\pi\alpha'eE[\,\delta(\sigma)-\delta(\pi-\sigma)\,]\right]\phi^\pm_m(\tau,\sigma)=\delta_{mn}\,{\rm sign } ( n \pm \epsilon)\nonumber
\qqq
where $\phi\stackrel{\leftrightarrow}{\partial_\tau}\psi=\phi\partial_\tau\psi-\psi\partial_\tau\phi$.
After a gauge transformation, the momentum density is 
\qq
\pi P_\pm=\partial_\tau X_\mp+i\pi\alpha'eE\,X_\mp(\delta(\sigma)-\delta(\pi-\sigma))\hskip0,3cm .
\qqq
Considering the usual canonical commutation relations
\qq
\left[X_{\mu}(\tau,\sigma),X_\nu(\tau, \sigma')\right]=0\hskip2,1cm\nonumber\\
\left[P_{\mu}(\tau,\sigma),P_\nu(\tau, \sigma')\right]=0\hskip2,2cm\\
\left[X_{\mu}(\tau,\sigma),P_\nu(\tau, \sigma')\right]=i\delta_{\mu\nu}\delta(\sigma-\sigma')\nonumber
\qqq
we find for $\mu,\nu=-,+$ the (gauge invariant) relations
\qq
\left[a_n^-, a_m^{+\dag}\right]=\delta_{nm}\hskip1,0cm\nonumber\\
\left[a_n^+, a_m^{-\dag}\right]=\delta_{nm}\hskip1,0cm\\
\left[\hat x^+, \hat x^-\right]=\frac{1}{4\alpha'eE}\hskip0,4cm .\nonumber
\qqq
Contrary to what we may think, $\hat x^\pm$ are not identified with the center-of-mass coordinate since the functions $\phi_n$ are not periodic in $\sigma$ and so they do not vanish by taking their mean-value integral.

 The open string Virasoro zero-mode restricted to the lightcone coordinates given by the energy-momentum tensor is
\qq
L_0^{(O,\, \ell)}\,=\,-\sum_{n=1}^\infty (n-i\epsilon)(a_n^+)^\ast a_n^-\,-\,\sum_{n=0}^\infty (n+i\epsilon)(a_n^-)^\ast a_n^+\,+\,a(\epsilon)
\qqq
with normal order constant $a(\epsilon)=i\epsilon(1-i\epsilon)/2$.
By a similar computation of the one-loop free energy $F$ as in Sect. 2.5, considering now four contributions from the unoriented open and closed strings (the annulus, M$\ddot{o}$bius strip, torus and Klein bottle), the rate of pair production per unit volume given by the imaginary part of $F$ is
\qq
\omega\,:=\,\Im (F)\,=\,\frac{1}{(2\pi)^{25}}\frac{\alpha'eE}{\epsilon}\sum_{i,k}(-1)^{k+1}\left(\frac{|\epsilon|}{k}\right)^{13}\,e^{-\pi k(M_i^2+\epsilon^2)/|\epsilon|}
\qqq
where the sum is taken over the physical states and the internal momentum.
We see that the rate production diverges as $\epsilon\rightarrow\infty$ corresponding to the upper bound on the electric field $E_{crit} = (2\pi\alpha' e)^{-1}$.

The dual model can be roughly seen by noting that the dynamics of the system is governed by the Born-Infeld action \cite{calkleb}. Under the change $2\pi\alpha' e E\leftrightarrow V$, this becomes the action for a relativistic point particle,
\qq
\CL_{BI}\propto \sqrt{1-(2\pi\alpha'E)^2}\hskip0.5cm \leftrightarrow\hskip0.5cm \CL_{particle}\propto \sqrt{1-V^2}\hskip0,3cm .
\qqq
The existence of a critical electric field is dual to the existence of a critical relativistic velocity.
This will become more explicit in the next section.


\section{The $T$-duality $E \leftrightarrow V$ on the $D$-brane action}

\hskip0,5cm The bosonic $D_{25}$-brane action with $A_\mu$ gauge fields on its world volume is given by  
\qq
S=\frac{1}{4\pi\alpha'}\int d\tau d\sigma\partial^a X^\mu\partial_a X_\mu
+ie\sum_{\mu=0}^{25}\int d\tau A^\mu(X^0,..., X^p)\partial_\tau X_\mu
\label{dbr25}
\qqq
where the gauge fields are restricted to depend only on the set of  $(X^0,...,X^p)$ coordinates and we have neglected the Kalb-Ramond $B_{\mu\nu}$ field contribution. Here all the string target space coordinates obey the Neumann boundary condition. Take the gauge $A^M=F^{M\alpha}X_\alpha$ for $\alpha=0,1,...,p$ and $M=p+1,...,25$. Considering only the disk-interactions, the low energy effective Lagrangian is given by the Born-Infeld Lagrangian 
\qq
\CL_{BI}\,\sim\,\sqrt{-{\rm det}(\eta_{\mu\nu}+2\pi\alpha' e F_{\mu\nu})}
\label{binf}
\qqq
where $\eta_{\mu\nu}$ is the metric on the brane. Performing a $T$-duality along the 
coordinates $X^M$, we obtain a $D_p$-brane describing some trajectory in the $D=26$ 
Minkowski spacetime given by $Y^M=2\pi\alpha' e A^M$. Under such transformation, 
the above Born-Infeld action (\ref{binf}) changes to the Nambu-Goto action of a 
relativistic boosted $D_p$-brane
\qq
\CL_{D_p}\,\sim\,\sqrt{-{\rm det}(\eta_{\alpha\beta}+\partial_\alpha Y^M\partial_\beta Y_M)}\hskip0,3cm .
\qqq
Here $Y^M$ is written in the gauge where the $p+1$ longitudinal coordinates parametrize the worldsheet of the brane. 

On the other hand, the sigma model describing the dynamics of a bosonic $D_p$-brane is given by 
\qq
S=\frac{1}{4\pi\alpha'}\int d\tau d\sigma\partial^a X^\mu\partial_a X_\mu
+\frac{1}{2\pi\alpha'}\sum_{M=p+1}^{25}\int d\tau Y^M(X^0,..., X^p)\partial_\sigma X_M
\label{dbrp}
\qqq
where $Y^M$ describes the trajectory of the $D_p$-brane and boundaries are taken to be the timelike surfaces $\sigma=0,\pi$.
From the actions (\ref{dbr25}) and (\ref{dbrp}) we may compute the two point correlation 
functions. The Neumman coordinates transform under T-duality to Dirichlet coordinates.
Both are related by
\qq
<\partial_\sigma X^i(\tau)\partial_\sigma X^j(\tau')>_{|_{Dirichlet}}=\hskip6,0cm\nonumber\\
\hskip2,0cm =-<\partial_\tau X^i(\tau)\partial_\tau X^j(\tau')>_{|_{Neumann}}=\frac{2\alpha' \delta^{ij}}{(\tau-\tau')^2}\hskip0,3cm .
\qqq
The dynamics of the brane is visualized by choosing a reference frame defined by a 
second $D_p$-brane in which the first brane is moving at velocity $V$. Open strings 
stretched between the two branes obey Neumann boundary conditions along $X^{1,...,p}$ 
coordinates and Dirichlet conditions on $X^{p+1,...,24}$ and 
\qq
X^{25}=\partial_\sigma X^0 = 0 \hskip1.0cm at \,\,\,\sigma=0\nonumber\\
X^{25}-vX^0=\partial_\sigma(X^0-vX^{25})=0 \hskip1.0cm at \,\,\,\sigma=\pi
\qqq
The solutions of these constraints are technically similar to closed string in a twisted sector of an orbifold with imaginary twist angle given by the velocity $e:={\rm arctanh}(V)$
\qq
X^0\pm X^{25} := X^\pm = i\sqrt \alpha '\sqrt{\frac{1\pm V}{1\mp V}}\times\hskip6.0cm\nonumber\\
\times \sum _{n=-\infty}^\infty\left[\frac{a_n}{\sqrt{n+ie}} e^{-i(n+ie)(\tau\pm\sigma)}
+\frac{\tilde a_n}{\sqrt{n-ie}} e^{-i(n-ie)(\tau\mp\sigma)}\right]
\qqq
where left-mode oscilators obey the canonical commutation relations
\qq
[a_n,a_m]=\delta_{n+m,0}
\qqq
with the same condition on the right-movers.
The existence of a relativistic limit in the electric field is now interpreted as an upper bound on the relativistic velocity, as was nicely explained in \cite{bachas}.

One may also look for the states associated to moving D-branes \cite{DiVecchia2}\cite{fpslr}. As we have seen in Sect.2.4, the boundary state associated to a static $D_p$-brane at position $y^i$ is given by
\qq
|B_X\rangle\,=\,(2\pi\sqrt{\alpha'})^{25-p}\delta^{25-p}(q^i-y^i)\,\exp\left(-\sum_{n=1}^\infty a_{-n}^\mu S_{\mu \nu} a_{-n}^\nu\right)|0\rangle|\tilde 0\rangle|p=0\rangle 
\qqq
where 
\qq
S_{\mu \nu}=(\eta_{\alpha \beta},-\delta_{ij})
\qqq
and we have chosen indices $\alpha\,,\,\beta$ ($i\,,j$) to label the components longitudinal (transverse) to the $D_p$-brane. 
The state associated to the boosted $D_p$-brane is characterized by 
\qq
|B,y,V\rangle = \exp^{iV_jJ^{0j}} |B,y(V)\rangle\nonumber\\
y^i(V)=y^i+V^i(V_jy^j)(\cosh|V|-1)/|V|^2
\qqq
where $J^{\mu \nu}$ is the boost operator
\qq
J^{\mu \nu}= q^\mu p^\nu - q^\nu p^\mu -i\sum_{n>0}\left(a_{-n}^\mu a_n^\nu - a_{-n}^\nu a_{n}^\mu +\tilde a_{-n}^\mu \tilde a_n^\nu - \tilde a_{-n}^\nu \tilde a_n^\mu \right)\hskip0,3cm .
\qqq
Considering the boost in a single transverse direction label as the $25-$ direction, 
the boosted state becomes
\qq
|B,y,V\rangle = (2\pi\sqrt{\alpha'})^{25-p}\sqrt{1-V^2}\delta(q^1- q^0V-y^1)\Pi_{i\neq 1}\delta(q^i - y^i)\times\hskip2.0cm\nonumber\\ 
\times\exp\left(-\sum_{n=1}^\infty a_{-n}^\mu \,{S_{\mu \nu}}_{\line(0,1){15}\mu\neq 0,25} a_{-n}^\nu\right)\exp\left(-(a_{-n}^0\,a_{-n}^1)M(V)^2
\left( \begin{array}{c}\tilde a_{-n}^0 \nonumber\\ \tilde a_{-n}^1 \end{array}\right)\right)|0\rangle|\tilde 0\rangle|p=0\rangle 
\qqq
with
\qq
M(V)=
\left( \begin{array}{c}
\,\,\cosh V \,\,\,\,\,\,\,-\sinh V \nonumber\\
-\sinh V \,\,\,\,\,\,\,\,\, \cosh V
\end{array} \right)
\qqq
where the notation $S_{\mu\nu}|_{\mu\neq 0,25}$ means that we neglect the directions $\mu=0,25$ of the target coordinates, longitudinal to the motion. Note that in the above state there is a  boost of left-movers relative to the right-movers, similar to what we have found for charged open strings immersed in a constant electric field \cite{calkleb}, see end of Sect.2.4.2.

We see from the above remarks that {\it a priori} the boost limit $V\rightarrow 1$ is non-smooth as expected to happen for any massive (tensive) object. This does not mean that in the moduli space there is no brane that may be interpreted as an infinitely boosted $D$-brane. So far, the dynamics of D-branes tells us that there is a relative boost of the left-movers with respect to the right-movers. We thus expect that a nullbrane should be described by a purely single chiral sector, left or right depending on its orientation in the spacetime. By 
the $V\leftrightarrow E$ duality, a formulation of a nullbrane will presumably give 
a picture of open strings with charged endpoints at a critical electric field.


\section{Space/space noncommutative field theory}

\hskip0,5cm To find the propagator for an open string immersed in a constant electromagnetic field in a target space with constant metric $g_{\mu\nu}$, we have to solve the following constraints for the Green's function 
\qq
D^{\mu\nu}(z,z')=<X^\mu(z)X^\nu(z')>\hskip3,0cm\nonumber\\
g_{\mu\nu}\partial\bar\partial\,D^{\mu\nu}(z,z')=-2\pi\alpha'\delta(z-z')\hskip2,6cm\nonumber\\
\left[g_{\mu\nu}(\partial-\bar\partial)\,+\,2\pi\alpha'F_{\mu\nu}(\partial+\bar\partial)\right]D^{\mu\nu}(z,z')\mid_{z=\bar z}=0
\label{prop}
\qqq
where $\partial=\partial/\partial z$, $\bar\partial=\partial/\partial\bar z$ and $\Im z\geq 0$. The solution is
\qq
D^{\mu\nu}(z,z')\,=\,-\frac{1}{2}\alpha'g_{\mu\nu}\ln|z-z'|^2\,+\,\frac{1}{2}\alpha'\left(\frac{g+2\pi\alpha'F}{g-2\pi\alpha'}\right)_{\mu\nu}\ln(z-\bar z')\,\nonumber\\
+\,\frac{1}{2}\alpha'\left(\frac{g-2\pi\alpha'F}{g+2\pi\alpha'F}\right)_{\mu\nu}\ln(\bar z-z')
\qqq
The propagator at the boundary $\sigma=\sigma'=0$ in the limit when $\tau\rightarrow \tau'$ is 
\qq
D_{\mu\nu}(\tau\rightarrow\tau')=-\alpha'\left[g+\frac{1}{2}\left(\frac{g-2\pi\alpha'F}{g+2\pi\alpha'F}\right)+\frac{1}{2}\left(\frac{g+2\pi\alpha'F}{g-2\pi\alpha'F}\right)\right]_{\mu\nu}\ln\Lambda\nonumber\\
=-2\alpha'\left[\frac{g}{(g-2\pi\alpha'F)(g+2\pi\alpha'F)}\right]_{\mu\nu}\ln\Lambda\hskip2,0cm
\qqq
where $\Lambda$ is a short distance cutoff. We may now find the 1-loop
 contribution to the perturbed system \cite{fr}
\qq
S_1=\frac{i}{2\pi}\int_{z=\bar z}ds\,\Gamma_\mu\partial_sX^\mu\hskip4,0cm\nonumber\\
\,\,\, =-\frac{i}{4\pi\alpha'}\int ds\,\partial_\nu F_{\mu\lambda}\partial_\tau X^\mu D^{\nu\lambda}(\tau,\tau')\mid_{\tau\rightarrow\tau'}
\qqq
and calculate the beta function, from which we get the equations of motion by imposing the fixed point condition
\qq
\beta_\mu=\lambda\frac{\partial}{\partial\lambda}\Gamma_\mu\hskip6,3cm\nonumber\\
=\partial^\nu F_\mu^{\,\,\,\,\lambda}\left[\frac{g}{(g-2\pi\alpha'F)(g+2\pi\alpha'F)}\right]_{\lambda\nu}=0\hskip0,3cm .
\qqq
They are also given by the effective Langrangian
\qq
L_{eff}\sim\sqrt{-{\rm det}(g_{\mu\nu}+2\pi\alpha'F_{\mu\nu})}\hskip0,3cm .
\qqq

A more general expression of the propagator at the boundary is \cite{seiwit}
\qq
\left< X^\mu(\tau)X^\nu(\tau')\right>\,=\,-\alpha'G^{\mu\nu}\log(\tau-\tau')^2\,+\,\frac{i}{2}\Theta_{\mu\nu}{\rm sign}(\tau-\tau')
\qqq
that should be compared with the two-point correlation function of the dissipative Hofstadter model, see end of Sect.2.6. Here the mobility and the Hall term are given by 
\qq
G^{\mu\nu}\,=\,\left(\frac{1}{g+2\pi\alpha'B}g\frac{1}{g-2\pi\alpha'B}\right)^{\mu\nu}\,\,,\hskip2,0cm\nonumber\\
\Theta^{\mu\nu}\,=\,-(2\pi\alpha')^2\left(\frac{1}{g+2\pi\alpha'B}B\frac{1}{g-2\pi\alpha'B}\right)^{\mu\nu}\hskip0,3cm .
\qqq
The term $G^{\mu\nu}$ is known in the string literature as the inverse effective metric seen by the open strings, that depends on the closed string metric $g_{\mu\nu}$. The Hall effect term is interpreted in condensed matter physics as the quantity that gives the commutation relations of operators from the short distance behaviour of time ordered product (the famous BJL relation, for a recent discussion see for example \cite{schomerus} and references therein) 
\qq
\left[X^\mu(\tau)\,,\,X^\nu(\tau')\right]\,=\,T\left(X^\mu(\tau)X^\nu(\tau^-)\,-\,X^\mu(\tau)X^\nu(\tau^+)\right)\,=\,i\Theta^{\mu\nu}
\qqq
that is interpreted as a spacetime with a noncommutative geometry. From the boundary 
correlation functions, we see that the spacelike coordinates of the open string ends 
become noncommutative in the allowed limit of infinite magnetic field 
$B_{ij}\rightarrow\infty$. This limit is the same as if we kept the magnetic field 
constant and take otherwise the field theory limit $\alpha'\rightarrow 0$, for which the string oscillators disappear. 


\section{Space/time noncommutativity:  Field Theory {\it vs} String Theory}

\hskip0,5cm We might think that the space/space noncommutativity in the presence of a magnetic field 
could be extended to its space/time counterpart by turning on an constant electric field, say 
along the $X$-direction
\qq
\Delta T\Delta X\sim \theta
\qqq
in the field theory limit. This is however not the case since there is a critical 
value of the electric field such that above it String Theory has no physical meaning. 
Thus it is not possible to take the field theory limit by considering the  
limit $E\rightarrow +\infty$. Nevertheless, it is possible to find a certain limit 
for which space/time noncomutativity arises in a new noncritical String Theory 
where open strings decouple from the closed strings, in particular from the 
gravitational sector of the theory \cite{sst}. Here, the open strings are confined 
to the branes, that is to say, they cannot interact with each other in the usual 
way to transform into a closed string leaving the brane and propagating freely in the 
bulk. In the new theory such a process would cost an infinite amount of energy. 
Charged open strings behave as a dipole oriented by the electric field lines. 
At the critical electric field limit, an infinite amount of energy would be necessary 
to push both ends of the string to meet.
Such a phenomenon is seen by a given worldsheet observer for which the
 boundary 
conditions are consistent with the modified light cone gauge
\qq
X^+=\tau+\tilde E\sigma 
\label{mlg}
\qqq
given in terms of the normalized electric field  $\tilde E = 2\pi\alpha' E$.

Note that for vanishing electric field and non-zero magnetic field, the gauge $X^+=\tau$ is perfectly consistent with the familiar boundary conditions on timelike boundaries. 
If we want to keep in mind that physics should be gauge invariant, we note that 
there is something peculiar in the consistency of boundary conditions with 
respect to  
timelike boundaries when we turn on the electric field. In the next Chapter we shall 
take the general expression for the Polyakov action that depends on the worldsheet 
metric, and we shall study the consistency of boundary conditions in the light-cone 
gauge $X^+=\tau$. We will see that the critical electric field is related to a Penrose 
limit in the worldsheet metric where boundaries become lightlike.

Nevertheless, it is not possible to find a field theory limit in the presence of 
the electric field. In fact, space/time noncommutative field theory is acausal 
and non-unitary, so we are glad not to have such a limit from  string theory.

To see this, we compute the $S$-matrix describing the scattering of two particles 
in 1+1 dimensions. Let us call $iM$ the non-trivial components of it 
\qq
\langle k_1, k_2|S|p_1, p_2\rangle=(2\pi)^2\delta^2(k_1+k_2-p_1-p_2)(1+iM)\hskip0,3cm .
\qqq
Consider the tree-level amplitude of $\phi^4$-theory with coupling constant $g$. At the center of mass frame, where the incoming and outgoing particles have the same absolute value of spatial momenta but with opposite directions, the non-trivial contribution to the $S$-matrix is                
\qq
iM\sim ig\hskip0,3cm .
\qqq
In the presence of space/time noncommutativity, the usual field multiplication
is replace by the $\star$-product with Moyal phase depending on the energy modes
\qq
[t,x]=i\theta\hskip6.0cm\nonumber\\
\phi_1\star\phi_2(x,t)\,=\,e^{i\frac{\theta}{2}(\partial_0^w\partial_1^z-\partial_1^w\partial_0^z)}\phi(w)\phi(z)\mid_{w=z=(x,t)}\hskip0,3cm .
\qqq
We now have 
\qq
iM\sim ig(\cos(4p^2\theta)+2)
\qqq 
where $p$ is the center of mass frame momenta of the incoming (and outgoing) particles. The presence of non-vanishing Moyal phase $\theta$ will produce an outgoing wave-function that is split into three parts, two of them concentrated at $x\sim \pm p_0\theta$ and 
the third one at $x=0$. The first two are interpreted as advanced and retarded waves, 
that is to say, they are created before and after the collision takes place. Since 
the Moyal phase depends on the energy of the particles, the bigger the energy, the 
bigger the time shifts. The problem comes from the advanced wave - if we want 
to preserve Lorentz invariance, it violates causality. The scattering particles 
resemble rigid rods that expand as their momentum increases, see \cite{sst2}. 

Fortunately, those pathologies are not present for space/time noncommutative string 
theory. Here the string oscillator modes play an essential role since they do not 
allow the creation of the pathological advanced wave packet. To see that, we start 
to compute the disk Veneziano amplitude without the presence of an electric field 
of four massless open strings. Using the Mandelstam variables
\qq
s=2p_1p_2\hskip0,3cm,\hskip0,6cm t=2p_1p_4\hskip0,3cm ,\hskip0,6cm u=2p_1p_3
\qqq
the disk amplitude splits into three terms
\qq
A_{D_2}(p_1,p_2,p_3,p_4)\sim g_s\delta(\sum_i p_i)\left[I(s,t)+I(t,u)+I(u,s)\right] 
\label{amp1}
\qqq
where $g_s$ is the open string coupling and the functions $I(x,y)$ are expressed in 
terms of the gamma functions. Consider now the case when an electric field is present. 
The new disk amplitude is obtained from the above one by replacing the metric 
$\eta_{\mu\nu}$ by the effective open string metric $G_{\mu\nu}$, $g_s$ by $G_s$ 
and by multiplying the terms by the Moyal phases. After proper modifications, 
the first term on (\ref{amp1}) is now given by
\qq
I^\theta(s,t)\sim G_s\left(K_{st}\exp^{2\pi i \tilde E s \ell_s^2}+\bar K_{st}\exp^{-2\pi i \tilde E s\ell_s^2}\right)\gamma(-2s\ell_s^2)\gamma(2s\ell_s^2)
\qqq
where we have expressed the kinematic terms involving the momenta $p_i$ by factors
 $K$ that in the noncommutative case are proportional to $s^2$. The second (advanced) 
phase is the one responsible for acausality in the field theory case. In the string case,
 the dependence on the gamma functions is crucial. If we expand their product 
into power series, the acausal behavior is cancelled and we are left with 
\qq
I^\theta(s,t)\sim G_s\,s\sum_{n>0\,odd}a_1 \exp^{2\pi i(n+\tilde E)s\ell_s^2}\,+\,a_2\exp^{2\pi i(n-\tilde E) s\ell_s^2}
\qqq
for some constants $a_1$ and $a_2$ independent of $s$. Let us remember that the new noncritical string theory was obtained by scaling the metric 
\qq
g^{-1}\sim 1-\tilde E^2\,\hskip2.0cm \ell_s^2=\alpha'/g
\qqq
where the noncommutative parameter in this limit is finite
\qq
\theta \sim 2\pi\alpha'\hskip0,3cm .
\qqq
By comparing the above string amplitude to the field theory case, we note that the string oscillators modify the noncommutative parameter by
\qq
\theta_n = 2\pi(n\pm\tilde E)\ell_s^2\,,\hskip1.0cm n>0\,,\,\,\,{\rm odd}
\hskip0,3cm .
\qqq 
Nevertheless, the string grows with energy, contrary to the intuitive Lorentz contraction. But this phenomenon is well known and explained in \cite{sussk1} as a consequence 
that at high energy, we are able to see more string-oscillating modes than at low 
energy and consequently, the string seems to grow with the energy. This is an 
important point of the black hole complementarity paradox.

We may interpret the phenomena as the occurence of a {\it phase transition} as we reach 
the critical limit for the electric field. It is possible that the phase transition 
gives a new unstable vacuum where the usual phase space is replaced by a 
noncommutative spacetime, i.e., the {\it phase transition} is a 
{\it space/phase-space transmutation} that might be associated to a new vacuum state. This raises the question of looking for a model where space/time noncommutativity arises 
without the presence of an electric field. In the next section we will see that the 
Schild action is one possible model \cite{yoneya}.


\section{Tensionless strings and the Schild action}

It is well known that the Polyakov action is classicaly equivalently to the Nambu-Goto action $S_{NG}$, which has the simple $n$-extension 
\qq
S_n=-\int d\tau d\sigma\, e\left[\frac{1}{e^n}\left(-\frac{1}{2\lambda^2}(\epsilon^{ab}\partial_aX^\mu\partial_bX^\nu)^2\right)^{n/2}+n-1\right]\nonumber\\
S_1=S_{NG}\hskip0,3cm . \hskip8,3cm
\qqq
The case $n=2$ was first proposed by Schild \cite{schild}. His action describes tensionless 
null strings travelling at the speed of light, including their endpoints for the 
case of open strings. We emphasize here that from the classical equations of motion 
for a free Nambu-Goto string, we see that the endpoints travel at the speed of light 
but the Nambu-Goto string should not be confused with the null Schild string, where 
 all the points travel at the speed of light and the string is tensionless. 

The quantization of the null string is given in the Appendix A where we show that 
the endpoint target coordinates describe a non-commutative spacetime geometry. 

On the other hand, the Polyakov action seems to be the unique two dimensional 
action which is tractable in the path-integral approach. Nevertheless, attempts 
are made for the case of the Schild action whose path-integral is of the form
\qq
Z=\int \CD \gamma\,\CD X\,\CD \psi\, e^{-S_{schild}}\hskip0,3cm .
\qqq
Here we have included the fermionic counterpart. The path-integral can be approximated by a quantum matrix IKKT-model \cite{ikkt}
\qq
Z=\sum_{n=0}^\infty \int dA\,d\psi\, e^{-S_{IKKT}}
\qqq
with 
\qq
S_{IKKT}=\alpha\left(-\frac{1}{4} {\rm Tr}[X_\mu,X_\nu]^2-\frac{1}{2}{\rm Tr}(\bar \psi \gamma^\mu[X_\mu,\psi])\right)+\beta\,{\rm Tr} 1
\qqq
where $X_\mu$ and $\psi$ are bosonic and fermionic $N\times N$ hermitian matrices respectively. In the large $N$ limit they correspond to the classical spacetime coordinates and the fermionic counterpart fields. The Schild action coincides with the $IKKT$ action in the 
large $N$ limit after we replace the commutator and the trace by the Poisson bracket 
and the integration
\qq
-i[\,,\,]\hskip0.7cm\Rightarrow\hskip0.7cm \{\,,\,\}_{_{PB}}\nonumber\\
{\rm Tr}\hskip0.5cm\Rightarrow\hskip0.5cm \int d^2\sigma\sqrt \gamma\hskip0,3cm .
\qqq 
Moreover, it follows that the Schild action is conformal invariant only if we impose the constraint on the Poison bracket
\qq
-\frac{1}{2}(\{X^\mu,X^\nu\})^2=\lambda^2
\qqq 
which is interpreted as a spacetime noncommutativity condition. After Dirac quantization, the spacetime coordinates are described by hermitian matrices and their physics by the IKKT-model. 


\chapter{Flat $H$-branes and time-dependent Conformal Field Theory}

\hskip0,5cm We have seen in the last Chapter that charged open strings in the presence 
of an electric field are related by T-duality to boosted $D$-branes, 
both bounded by a certain critical value. In the electric case, because of the 
pair-production, we require to spend an infinite amount of energy to reach the 
critical value on the electric field. In the $T$-dual picture this phenomenon 
is interpreted as an infinite amount of energy necessary to infinitely boost 
a tensive D-brane. Nevertheless, in the worldsheet sense, the behaviour of the 
theory exactly at the critical value seems to present no pathology - it would 
describe some sort of a nullbrane. 

The picture we have in mind is that in the critcal limit there is a space/phase-space 
transmutation, that is to say, the usual momenta-coordinate commutation relations 
are replaced by the canonical commutation relations of a noncommutative spacetime 
vacuum. At this phase transition, infinitely boosted $D$-branes are replaced in 
the moduli space by $H$-branes, described by chiral and squeezed Ishibashi states
\qq
D-brane\,\, \longrightarrow\,\, H-brane\hskip0,3cm .
\qqq
Note that to describe a nullbrane in terms of boundary conditions, we need to have 
a Neumann condition for, say, $X^+$ and a Dirichlet condition for $X^-$, 
where $X^\pm$ are target lightcone coordinates. Is this possible?

To answer this let us recall how one gets boundary conditions for the bosonic 
open strings (for more details see \cite{pl}). The starting point is the variation 
of the Polyakov action that gives equations of motion from the bulk terms plus boundary conditions from the boundary term
\qq
\delta S_P=-\frac{1}{4\pi\alpha'}\int_\Sigma d\tau d\sigma \delta X^\mu\partial_a\left((-\gamma)^{1/2}\gamma^{ab}\partial_b X_\mu\right)\,-\,\frac{1}{4\pi\alpha'}\int_{\partial \Sigma} ds \delta X^\mu t^a\partial_a X_\mu
\qqq
where $t^a$ is a unit tangent vector to the boundary. Take the worldsheet metric in the conformal gauge and the familiar timelike worldsheet boundaries. The boundary conditions on the null and transverse target coordinates are
\qq
\delta X^+\partial_\sigma X^- = 0\nonumber\\
\delta X^-\partial_\sigma X^+ = 0\nonumber\\ 
\delta X^i\partial_\sigma X^i = 0
\label{3bc}
\qqq
where $i=2,3,...,25$ labels the transverse coordinates. From the first line of (\ref{3bc}), we see that if we choose the Dirichlet condition for say $X^+$, then 
$\partial_\sigma X^+\neq 0$. Then from the second line we are forced to choose also a Dirichlet condition for $X^-$, since $X^+$ do not obey the Neumann condition. 
By the same argument, if we choose the Neumann condition for $X^-$, then we will 
end also with the Neumann condition for $X^+$. On the transverse coordinates there 
are no such restrictions since boundary conditions do not have a cross product 
as in $+,-$ case. We can choose for each transverse coordinates, either Dirichlet 
or Neumann conditions. It seems that a $D$-brane with a single null direction 
is impossible.

Is this really so?


\section{Deformation of worldsheet boundaries}

\hskip0,5cm We start to analyze time-dependent conformal field theory considering boundary perturbations as a source of a dynamical deformation of the usual timelike worldsheet boundaries $\sigma=\rm{const}$ \cite{KR}. Consider the string worldsheet $\Sigma$ parametrized by the coordinates $(\tau,\sigma)$ of signature $(-,+)$, with $\tau$ the dynamical variable and boundaries defined at $\sigma=0,\ell$. We take the non-linear sigma model describing charged open string immersed in a constant (normalized) electric field $0\leq \tilde E\leq 1$ along the $(T,X)$ spacetime plane \footnote{In \cite{KR} we have conjectured the existence of a boundary term that was added to the Polyakov action with a $\beta$ parameter taking values between $0$ and $1$. We have later identified the physics of such conjectured model with the present one where $\beta=\tilde E$.}  
\qq
S_{\tilde E}^{open}\,=\,-\frac{1}{4\pi\alpha'}\int_\Sigma d\tau d\sigma (-\gamma)^{1/2} \gamma^{ab} \partial_{a}X^{\mu} \partial_{b}X_{\mu}\,
-\,\frac{\tilde E}{4\pi\alpha'}\int_{\partial\Sigma} ds X^{+}t^a\partial_a X^{-}
\qqq
where $t^a$ is a unit vector tangent to the boundary and we have defined $X^\pm:=(T\pm X)/\sqrt 2$. Now we dynamically construct the string worldsheet metric as in \cite{pl} by working in the lightcone gauge $X^+=\tau$ with two more conditions on the worldsheet metric
\qq
\partial_\sigma\gamma_{\sigma \sigma}=0\hskip1.0cm {\rm det}(\gamma_{ab})=-1\hskip0,3cm .
\qqq
Note that the lightcone gauge is different from (\ref{mlg}) and this is the cause for the dynamical deformation of worldsheet boundaries from the familiar timelike to lightlike one.

From the variational principle for the above action, the boundary condition for $X^+=\tau$ target coordinate gives the following restriction on the worldsheet metric \cite{KR}
\qq
ds^2\,=\,\gamma_{\tau \tau}d\tau^2\,+\,\gamma_{\sigma\sigma}d\sigma^2\,-\,2\tilde E\,d\tau d\sigma
\label{ds}
\qqq
with
\qq
\gamma_{\sigma\sigma}\gamma_{\tau\tau}\,=\,\tilde E^2 -1
\label{v2}
\qqq
where
\qq
\gamma_{\sigma\sigma}=\gamma_{\sigma\sigma}(\tau)\hskip0,3cm .
\qqq
The complete set of boundary conditions are
\qq
\partial_\sigma X^+\,=\,0\hskip1.6cm\nonumber\\
2\tilde E\partial_\tau X^-\,=\,f(\tilde E)\partial_\sigma X^-\nonumber\\
\tilde E\partial_\tau X^i\,=\,f(\tilde E)\partial_\sigma X^i\hskip0.2cm
\label{bc}
\qqq
where $f(\tilde E)=-\gamma_{\tau\tau}(\tau)$ and we impose that $f(0)=1$ and $f(1)=0$, consistently 
with ($\ref{v2}$). In general we have a mixing of Dirichlet and Neumann conditions as expected. Nevertheless, for $\tilde E\rightarrow\tilde E_{crit}=1$ there is a clear separation of boundary conditions on the target lightcone coordinates
\qq
\partial_\sigma X^+=0\hskip0,3cm,\hskip0.6cm\partial_\tau X^-=0\hskip0,3cm,\hskip0.6cm\partial_\tau X^i=0 
\qqq
at $\sigma=0,\ell$. This is the technical starting point to define a nullbrane!
\vskip0.3cm

From the non-linear sigma model action we define its Lagrangian by considering $\tau$
as the dynamical variable. The Hamiltonian in lightcone gauge is \cite{KR}
\qq
H_{\tilde E}^{open}\,=\,\frac{\ell}{4\pi\alpha'p^{+}}\int_{0}^{\ell}d\sigma \left[2\pi\alpha' \Pi^{i}\Pi^{i}+\frac{1}{2\pi\alpha'}\partial_\sigma X^{i}\partial_\sigma X^{i}+2\tilde E\Pi^{i}\partial_\sigma X^{i} \right]\nonumber\\
-\frac{\tilde E}{2\pi\alpha'}\int_{0}^{\ell}d\sigma \partial_\sigma Y^{-}-{f(\tilde E)\over 4 \pi \alpha '}\left[X^+\partial_\sigma Y^-\mid_{\sigma=0} + X^+\partial_\sigma Y^-\mid_{\sigma=\ell}\right] 
\qqq
 where we have split $X^-(\tau,\sigma)=Y^-(\tau, \sigma)+\bar{X}^-(\tau)$ with
\qq
\bar{X}^-(\tau)={1\over \ell}\int_0^\ell d\sigma X^-(\tau, \sigma)\hskip0,3cm .
\qqq
The momentum conjugate to $\bar{X}^-$ is
\qq
p_-=-p^+=-{\ell \over 2\pi \alpha '}\gamma_{\sigma\sigma}(\tau)
\qqq
and for $X^i$ is
\qq
\Pi^i={p^+\over \ell}\partial_\tau X^i - {\tilde E \over 2\pi \alpha '}\partial_\sigma X^i\hskip0,3cm .
\qqq
The equations of motion are 
\qq
\partial_\tau p^+\,=\,0
\nonumber\\
-{1\over c}\partial_\tau^2X^i-2\tilde E\partial_\tau\partial_\sigma X^i+c(1-\tilde E^2)\partial_\sigma^2X^i=0
\label{eq1}
\qqq
for $c={\ell \over 2 \pi \alpha ' p^+}$ where $p^+$ is a conserved quantity. The equation of motion (\ref{eq1}) becomes the wave equation \qq \gamma^{ab}\partial_a\partial_bX=-\gamma_{\sigma\sigma}\partial_\tau^2X+2\gamma_{\tau\sigma}\partial_\tau\partial_\sigma X-\gamma_{\tau\tau}\partial_\sigma^2X=0\qqq if we take
\qq
\gamma_{\sigma\sigma}=1/c\hskip2.0cm
\gamma_{\tau\tau}=-c(1-\tilde E^2)\,,
\qqq
see (\ref{ds}). Moreover, it's convenient to take the combination $\ell/p^+$ such that we have $c\sqrt{1-\tilde E^2}=1$ and $f(\tilde E)=-\gamma_{\tau\tau}=\sqrt{1-\tilde E^2}$, so that in particular the critical limit takes the form 
\qq
c\rightarrow \infty \hskip0.5cm {\rm when}\hskip0.5cm \tilde E\rightarrow 1 \hskip0.5cm {\rm such\,\,\, that}\hskip0.5cm c(1-\tilde E^2)\rightarrow 0\hskip0,3cm .
\qqq

We see that by turning on a boundary perturbation given by the $U(1)$-currents $(\partial_\tau X^-,\partial_\tau X^+)$, we have dynamically deformed the familiar timelike boundaries $\sigma=\rm{const}$ to a new form that can be read from the perturbed worldsheet metric in Lorentz signature
\qq
ds^2=-\sqrt{1-\tilde E^2}d\tau^2-2\tilde E d\tau d\sigma+\sqrt{1-\tilde E^2}d\sigma^2
\qqq
where $\tau$ is the dynamical variable. 
\vskip0.3cm
At the critical limit, the equations of motion are 
\qq
\gamma^{ab}\partial_a\partial_b X=\partial_\tau\partial_\sigma X=0
\qqq
with general solution given as usual by a linear combination of left- and right-chiral movers
\qq
X=X_L(\tau)+X_R(\sigma)
\qqq
with the worldsheet coordinates $\tau$ and $\sigma$ parametrizing the left- and right-chiral movers. So they are nullcoordinates, consistent with (\ref{ds}) where we have at the critical limit $ds^2=d\tau d\sigma$ and the worldsheet boundaries $\sigma = \rm{const}$ become null.
By a careful analysis, it is interesting to note that the Hamiltonian takes the unusual form \cite{KR} 
\qq
H^{open}_{1}=-\frac{1}{2\pi\alpha'}\int_0^\ell d\sigma \partial_\sigma Y^-
\qqq
with transverse conjugate momenta 
\qq
\Pi^i=-\frac{1}{2\pi\alpha'}\partial_\sigma X^i\hskip0,3cm .
\qqq
Note that all transverse oscillators are frozen in time.
\vskip0.3cm
Similarly, at the critical limit we may add the fermionic contribution to the system given by the action 
\qq
S^{open}_1(\psi)={1\over 8\pi}\int_0^\ell d\sigma \int_{-\infty}^\infty d\tau 
\,(\psi^\mu_L\partial_\tau\psi_{\mu L} + \psi^\mu_R\partial_\sigma\psi_{\mu R}) 
\label{ferac}
\qqq
where $\psi_L,\,\psi_R$ are the left- and right-Majorana spinors. The momenta conjugate to the fermionic fields are
\qq
\Pi_L^\mu={\delta L\over \delta (\partial_\tau\psi_{\mu L})}={1\over 8\pi}\psi^\mu_L\nonumber\\
\Pi_R^\mu={\delta L\over \delta (\partial_\tau\psi_{\mu R})}=0\hskip0,8cm
\qqq
and the fermionic contribution to the Hamiltonian is
\qq
H^{open}_1(\psi)=-{1\over 8\pi}\int d\tau d\sigma \,\psi^\mu_R\partial_\sigma\psi_{\mu R} \hskip0,3cm .
\qqq


\section{Perturbative analyses of $H$-branes in the open string channel}

\hskip0,5cm Let us define the the variables $\xi^\pm_\tau=\delta_\tau X^\pm, \xi^\pm_\sigma=\delta_\sigma X^\pm$ where $ X^\pm$ are the null target coordinates. The perturbative action at the critical electric field $\tilde E=1$ is
\qq
-\frac{1}{2\pi\alpha'}\int_\Sigma \CL\,=\,-\frac{1}{2\pi\alpha'}\int_\Sigma d\tau d\sigma\, \xi_\tau^-\,\xi_\sigma^+
\qqq
with $\Sigma=\{(\tau, \sigma)\mid ds^2=2d\tau d\sigma\}$ with boundary $\delta\Sigma=\{(\tau,\sigma)\mid \sigma=0,\pi\}$. We use the first order formalism to quantize the system. The one-form is
\qq
\alpha\,=\,-\xi_\tau^-\xi_\sigma^+ d\tau\wedge d\sigma\,+\,\xi_\sigma^+ d X^-\wedge d\sigma\,+\,\xi_\tau^- d\tau\wedge d X^+\hskip0,3cm .
\qqq
The equations of motion are 
\qq
 X^\star(i_{\delta X^\pm}d\alpha)=0\,\Leftrightarrow \partial_\sigma\partial_\tau X^\pm\,=\,0
\qqq 
with boundary conditions
\qq
 X^\star(i_{\delta X^+}\alpha)=0\,\Leftrightarrow\, \xi^-_\tau=0 \hskip1.0cm {\rm at}\,\,\,\, \partial\Sigma\hskip0,3cm .
\qqq
The space of solutions are given in terms of left- and right-sectors
\qq
 X^+(\tau,\sigma)= X_R^+(\sigma)\,+\, X_L^+(\tau)\,+\, X^+_0\nonumber\\
 X^-(\tau,\sigma)= X_R^-(\sigma)\,+\, X^-_0\hskip1,8cm
\qqq
with 
\qq
\Omega\,=\,\int_0^\pi\partial_\sigma(\delta  X^+)\wedge\delta X_R^- d\sigma
\qqq
the symplectic two-form of the phase-space. The degeneracy of this form is given by the left-sector of the $ X^+$ target variable plus its zero mode. They are gauged away from the space of solutions
\qq
 X_R^+(\sigma)\,+\, X_L^+(\tau)\,+\, X^+_0\,\longrightarrow\,  X_R^+(\sigma)
\qqq
so that they become non-dynamical, i.e. all string oscillators are frozen in time.

We proceed with the quantization of the system by looking at the Poisson brackets. Given a certain function of the phase space $\CF$ with variation 
\qq
\delta \CF\,=\,\int_0^\pi f^-(\sigma)\delta X^+ d\sigma\,+\,\int_0^\pi f^+(\sigma)\delta X_R^- d\sigma
\qqq
we define the associated Hamiltonian vector field 
\qq
\CX_\CF\,=\,\int_0^\pi g^-(\sigma)\frac{\delta}{\delta  X_R^-}d\sigma\,+\,\int_0^\pi g^+(\sigma)\frac{\delta}{\delta X^+}d\sigma
\qqq
using the phase space symplectic structure $\delta\CF\,=\,i_{\CX_\CF}\Omega$ giving the relations
\qq
f^+(\sigma)\,=\,\partial_\sigma g^+(\sigma)\hskip4,7cm\nonumber\\
f^-(\sigma)\,=\,\partial_\sigma g^-(\sigma)\hskip0.3cm ,\hskip0,6cm g^-(0)\,=\,g^-(\pi)\,=\,0 
\label{i1}
\qqq
so that in particular 
\qq
\int_0^\pi f^-(\sigma)d\sigma\,=\,0\hskip0,3cm .
\qqq
From the Poisson bracket $\{\CF, \CF'\}_{PB}=\CX_\CF(\CF')$ we find the relations
\qq
\{ X_R^-(\sigma)\,,\, X^+(\sigma', \tau)\}_{PB}\,=\,\Theta(\sigma-\sigma')\nonumber\\
 \{ X^+(\sigma, \tau)\,,\, X_R^-(\sigma')\}_{PB}\,=\,\Theta(\sigma-\sigma')
\qqq
with $\Theta(\sigma)$ the Heaviside function. Note that both relations are compatible with each other because the phase space symplectic structure has a degeneracy so that the $ X^+$ coordinate is only defined up to a $\tau$-dependent function.

We can see that the complete set of mode oscillators satisfying $(\ref{i1})$ is 
\qq
 X^-_m\,=\,\sqrt{2/\pi}\int_0^\pi \sin (m\sigma)\,  X_R^-(\sigma)\,d\sigma\nonumber\\
 X^+_m\,=\,\sqrt{2/\pi}\int_0^\pi \cos (m\sigma) \, X_R^+(\sigma)\,d\sigma
\qqq
for $m > 0$. They are respectively a set of Dirichlet and Neumann modes where the Neumann zero mode is gauged away. The canonical commutation relations are
\qq
\{ X_m^-, X_n^+\}=\frac{\delta_{m,n}}{m}\,.
\qqq
They should be interpreted as non-commutation relations among the null target coordinates. 
Since these oscillator modes behave as an $(x,p)$ set, we can define the creation and annihilation set as $(x+ip, x-ip)$ by
\qq
{1\over n}\alpha_n={1\over \sqrt 2}( X^-_n+i X^+_n)\hskip0.3cm,\hskip0,6cm {1\over n}\alpha_n^\star={1\over n}\alpha_{-n}={1\over \sqrt 2}( X^-_n-i X^+_n)
\qqq
for $n>0$ and  
\qq
[\alpha_n,\alpha_{-n}]=in
\qqq
for which we define the Virasoro zero mode
\qq
L_0=\sum_{n\neq 0}\alpha_{-n}\alpha_n\hskip0,3cm .
\qqq


\section{Majorana spinors, ghost, superghost and the null boundaries}

\hskip0,5cm We now generalize the previous analysis of null worldsheet boundaries to the case of superstrings, where the fermionic contribution is given by ($\ref{ferac}$) and we continue to restrict ourselves to lightcone target fields. Since the new type of boundaries will not modify the familiar equations of motion, we continue to expand the Majorana spinors as
\qq
\psi^\pm_L(\tau)=\sum_{r\in\NZ + \nu}\psi_r^\pm e^{-ir\tau}\nonumber\\
\psi^\pm_R(\sigma)=\sum_{r\in\NZ + \tilde\nu}\tilde\psi_r^\pm e^{-ir\sigma}
\qqq
where $\nu$ and $\tilde\nu$ can take the values $0$ (Ramond sector), or $1/2$ (Neveu-Schwarz sector). From the boundary contribution of the variation principle we have
\qq
\psi_R^-\,\delta\psi_R^+\,=\,\psi_R^+\,\delta\psi_R^-\,=\,0
\qqq
at $\sigma=0,\pi$. It is sufficient to consider for example
\qq
\psi_R^-(\tau,0)\,=\,\psi_R^-(\tau,\pi)\,=\,0
\qqq
without any boundary condition on $\Psi_R^+$. Since $\psi^-_R=\psi^-_R(\sigma)$ we have
\qq
\tilde\psi^-_r = 0\hskip1.0cm\forall r\in \NZ+\tilde\nu\hskip0,3cm .
\qqq 
Nevertheless, we could have imposed the same boundary conditions on fermionic field $\psi_R^+$ 
\qq
\tilde\psi_r^+ = 0 \hskip1.0cm\forall r\in \NZ+\tilde\nu\hskip0,3cm .
\qqq
Note that if we had taken the null boundaries at $\tau=0,\pi$, we would have the previous conditions on opposite chiral $\psi_L^\pm$ fields
\qq
\psi^-_r=0\hskip1.0cm {\rm or} \hskip1.0cm \psi^+_r=0\hskip0.5cm \forall r\in \NZ+\nu\hskip0,3cm .
\label{mbc}
\qqq
Moreover, we would like to have at the null boundaries $\sigma=0,\pi$ the BRST invariance given by the condition on the BRST charges
\qq
Q+\tilde Q =0
\label{brst}
\qqq
where 
\qq
Q=\sum_n:(L_{-n}^{(\alpha,\psi)}c_n+G_{-n}^{(\alpha,\psi)}\gamma):\nonumber\\
-{1\over 2}\sum_{m,n}(m-n):c_{-m}c_{-n}b_{m+n}:\nonumber\\
+\sum_{m,n}({3\over 2}n+m):c_{-n}\beta_{-m}\gamma_{m+n}:\nonumber\\
-\sum_{m,n}:\gamma_{-m}\gamma_{-n}b_{m+n}:-ac_0
\qqq
with a similar expression for the right chiral fields. As we have imposed the boundary conditions on the bosonic and Majorana fields
\qq
\tilde\alpha_n^+=\tilde\psi_{n+\tilde\nu}^+=0\,\,\,\,\forall n
\qqq
we see (so far we are neglecting transverse coordinates to the spacetime lightcone) $\tilde L_m^{(\alpha, \psi)}=\tilde G_m^{(\alpha, \psi)}=0$ so that for condition $(\ref{brst})$ to hold, we need to impose the boundary conditions on the ghosts and superghosts
\qq
c_n=\gamma_n=0\nonumber\\
\tilde b_n=\tilde \beta_n=0\,\,\,\,\,\forall n
\label{gbc}
\qqq
at $x^+=0,\pi$. 
Taking a variation on the Faddeev-Popov action 
\qq
S_{FP}={1\over \pi}\int d\tau \int_0^\pi d\sigma (c^-\partial_\sigma b_{--}+c^+\partial_\tau b_{++})
\qqq
the null boundary conditions
\qq
c^-\delta b_{--}=0
\qqq
are in fact satisfied by considering only $c^-$ and we are free to set $b_{++}=0$ as we have in $(\ref{gbc})$.


\section{Chiral closed strings and null boundaries}

\hskip0,5cm Our main qualitative result is easy to understand. The usual mixing boundary conditions for charged open strings immersed in a constant electric field
\qq
\sqrt{1-\tilde E^2}\,\partial_\sigma X-\tilde E\partial_\tau X=0\hskip1.0cm \sigma\,=\,0,\,\pi
\qqq
are in the critical limit $\tilde E_{crit}=1\,$ rewritten as a single boundary condition
\qq
\partial_\tau X=0\hskip1.0cm \sigma\,=\,0,\,\pi
\qqq
so that we have interpolated the Neumann condition for $\tilde E=0$ to a Dirichlet condition for $\tilde E=1$. On the other hand, the induction of null boundaries on the worldsheet 
allows us to interpret this condition as a Dirichlet condition with respect to the null boundary, i.e., a $(D;n)$-condition in time-dependent string theory.

It is easy to visualize that the no-flux of energy condition through the null boundary is  
$\tilde T(\tau)=0$, so that the boundary states must satisfy the condition 
\qq
\tilde L_n^{(C;\ell)} |B\rangle = 0 \hskip0,3cm .
\qqq
The restriction that we had on the central charge $\,c=\tilde c\,$ by requiring $Diff(S^1)$ modular invariance of the boundary states, is now replaced by the condition $\,\tilde c=0\,$. Such a condition has appeared in the past few years in the study of LCFT, see \cite{shinsuke} for a review. The second consequence of null boundaries is that clearly there is no coupling 
between left- and right-chiral oscillator movers. In the target space, the defect caused 
by the boundary is interpreted as some kind of infinitely boosted D-brane; an H-brane.

We now show that it is possible to have purely Ishibashi chiral states along the target null direction without introducing singularities in quantities that describe the physical system, as for example, the (general) tree-level closed string amplitudes. Neglecting the zero-modes, note that the closed string left-modes on the null direction obey the Heisenberg algebra
\qq
\left[\alpha^+_m,\alpha^-_{-n}\right]=m\delta_{mn}\hskip0,4cm\nonumber\\
\left[\alpha^+_{-m},\alpha^-_n\right]=-m\delta_{mn}
\qqq
for $m , n \geq 1$. The chiral lightcone Heisenberg algebra is of the same form as if we had considered left- and right-chiral Heisenberg algebras for a given transverse coordinate
\qq
\left[\alpha_m,\alpha_{-n}\right]=m\delta_{mn}\hskip0,4cm\nonumber\\
\left[\tilde \alpha_{-m},\tilde \alpha_n\right]=-m\delta_{mn}
\qqq
for $m , n \geq 1$. As both algebras are indistinguishable, we can make the next identifications by the one-to-one maps
\qq
\alpha_m\leftrightarrow\alpha^+_m\nonumber\\
\alpha_{-m}\leftrightarrow\alpha^-_{-m}\nonumber\\
\tilde\alpha_m\leftrightarrow\alpha^-_m\nonumber\\
\tilde\alpha_{-m}\leftrightarrow\alpha^+_{-m}
\qqq
for $m>0$.
We see that the familiar non-chiral and coherent Ishibashi state of a D-brane
\qq
|D\rangle=\exp\left(\sum_{m\geq 1}{1\over m}\alpha_{-m}\tilde \alpha_{-m}\right)|0\rangle
\qqq
is mapped to the chiral and coherent Ishibashi state along a null direction
\qq
|N\rangle=\exp\left(\sum_{m\geq 1}{1\over m}\alpha_{-m}^- \alpha_{-m}^+\right)|0\rangle
\qqq
as well as 
\qq
\langle D|=\langle 0|\exp\left(\sum_{m\geq 1}{1\over m}\alpha_m\tilde \alpha_m\right)
\qqq
to 
\qq
\langle N|=\langle 0|\exp\left(\sum_{m\geq 1}{1\over m}\alpha_m^+ \alpha_m^-\right)\hskip0,3cm .
\qqq
We proceed to the analysis of the Majorana spinors where for $m,n >0$ the anticommutation relations for a single coordinate
\qq
\{\psi_r, \psi_{-t}\}=-\{\tilde \psi_{-r},\tilde\psi_t\}=\delta_{rt}
\qqq
obeys the same algebra as for Majorana spinors in lightcone coordinates
\qq
\{\psi^+_r, \psi^-_{-t}\}=-\{\psi^+_{-r},\psi^-_t\}=\delta_{rt}
\qqq
for $r,t>0$. We thus extend the previous map to
\qq
\psi_r\leftrightarrow\psi_r^+\nonumber\\
\psi_{-r}\leftrightarrow\psi_{-r}^-\nonumber\\
\tilde\psi_r\leftrightarrow\psi_r^-\nonumber\\
\tilde\psi_{-r}\leftrightarrow\psi_{-r}^+
\qqq
for $r>0$. 
The fermionic D-Ishibashi state 
\qq
|D\rangle_\psi=\exp\left(\pm i\sum_{r>0}\psi_{-r}\tilde\psi_{-r}\right)|0\rangle
\qqq
is mapped to a fermionic N-Ishibashi state
\qq
|N\rangle_\psi=\exp\left(\pm i\sum_{r>0}\psi^-_{-r}\psi^+_{-r}\right)|0\rangle
\qqq
with analogous relations for $\langle D|_\psi$ and $\langle N|_\psi$.
Moreover, we also see that the non-chiral closed string Hamiltonian restricted to a single coordinate
\qq
H^{(C)}=L_0^{(C)}+\tilde L^{(C)}_0 =\hskip7,5cm\nonumber\\
=\sum_{n\geq 1}(\alpha_{-n}\alpha_n+\tilde \alpha_{-n}\tilde \alpha_{n})\,+\,\sum_{r\in\NZ^+-\nu}r\,\psi_{-r}\psi_r\,+\,\sum_{r\in\NZ^+-\tilde\nu}r\,\tilde\psi_{-r}\tilde\psi_r
\qqq
is mapped to the chiral contribution of a closed string Hamiltonian restricted to the lightcone coordinates 
\qq
L_0^{(C;\ell)}=\sum_{n\geq 1}(\alpha^-_{-n}\alpha_n^++\alpha_{-n}^+\alpha_n^-)+\sum_{r\in\NZ^+-\nu}r\,(\psi_{-r}^-\psi_r^++\psi_{-r}^+\psi_r^-)
\qqq
where we have imposed $\nu=\tilde\nu$.

In the Cardy program we have seen by worldsheet duality that $D$-branes are naturally coupled to non-chiral closed strings, see (\ref{card}), and that the closed string amplitude is finite.
Nevertheless, we have the straightforward relation
\qq
\langle D_\beta|\tilde q ^{{1\over 2}(L^{(C)}_0+\tilde L^{(C)}_0)}|D_\alpha\rangle
=\langle N_\beta|\tilde q ^{{1\over 2}L_0^{(C;\ell)}}|N_\alpha\rangle
\qqq
since the algebras on both hand sides are the same by the previous identifications. 

In this simple example we see that there is enough structure on the lightcone mode oscillators so that we may disregard one of the two possible chiral sectors after imposing null boundaries, without introducing singularities in the physical system. Despite the fact that the chiral 
coherent states might seem good candidates to describe our $H$-branes, we will later
see that the latter are instead described by squeezed chiral states. See \cite{schumi} 
for a review of squeezed states and their applications to optics.


\section{$H$-branes and chiral squeezed Ishibashi states}

\hskip0,5cm For the calculation of the Ishibashi state corresponding to the $(D;n)$-boundary condition, we use in particular the one-dimensional path integral approach of \cite{callan} 
for the boundary state formalism. We consider a one-loop open string diagram by making 
a periodic identification on the variable $\sigma\sim \sigma+2\pi T$. At the endpoints 
we may impose the conditions
\qq
(D;n)\,:\, \partial_\tau X=0\hskip2.0cm {\rm open}\,\,\,\,\nonumber\\
(N;n)\,:\, \partial_\sigma X=0\hskip2.0cm {\rm string}
\qqq
at $\sigma=0,\pi$. By interchange $\tau\leftrightarrow \sigma$, the one-loop diagram is now a tree-level closed string channel. After we rescale the periodicity of $\sigma$ to $\sigma\sim \sigma+2\pi$ we have respectively
\qq
(D;n)\,:\, \partial_\sigma X=0\hskip2.0cm {\rm closed}\nonumber\\
(N;n)\,:\, \partial_\tau X=0\hskip2.0cm {\rm string}
\qqq
at $\tau=0,\pi/T$. Note that the interchange $\tau\leftrightarrow \sigma$ comes naturally associated to the interchange of left-movers with right-movers
\qq
\tau\leftrightarrow \sigma\hskip1.0cm\Longleftrightarrow\hskip1.0cm\alpha_n\leftrightarrow\tilde\alpha_n\hskip0,3cm .
\label{LtoR}
\qqq

We start to examine the case of closed string $(D;n)$-boundary condition for the coordinate $X=X^-$. In general we might think that the $N$-brane emits a closed string of the form $X^-(\tau,\sigma)=X^-(\tau)+X^-(\sigma)$ a general solution of the equations of motion $\partial_\tau\partial_\sigma X^\pm=0$. By imposing the closed $(D;n)$-boundary condition we see in fact that the $H$-brane has created a {\bf chiral closed string} $X^-=X^-(\tau)$ that will be annihilated by the second $H$-brane
\qq
X^-(\tau)=\hat q^-+\frac{\alpha^-_0}{2}\tau\,+\,{i\over 2}\sum_{n\neq0}{\alpha^-_n\over n}e^{-in\tau}\nonumber\\
=\hat q^-+{\alpha^-_0\over 2}\tau\,+\,{1\over 2}\sum_{n>0}\frac{X^-_n}{\sqrt n}\hskip0,8cm
\qqq
where  
\qq
 X^-_n\,=\,a^-_n e^{-in\tau}+a_n^{-\dag} e^{in\tau}
\label{1}
\qqq
for $\tau=0,\,\,\pi/T$. We have used the redefinitions $a^-_n={i\over \sqrt{n}}\alpha^-_{n}$ and $a_n^{-\dag}={-i\over \sqrt{n}}\alpha^-_{-n}$ with $n>0$ that for the closed string channel obey the usual canonical commutation relations
\qq
[a_m^+,a_n^-]=\delta_{m+n,0}
\qqq 
for $m\neq 0$. 
Given the single set (\ref{1}) of bosonic coordinates, we interpret it as a eigenvalue condition for a given eigenstate
\qq
(a^-_ne^{-in\tau}+a_n^{-\dag} e^{in\tau}- X^-_n)|X^-\rangle=0\hskip0.5cm\forall n>0\hskip0,3cm .
\qqq
A possible solution is 
\qq
|X^-\rangle = \prod_{n=1}^\infty |X^-_n\rangle\hskip8,0cm\nonumber\\
|X^-_n\rangle = (2\pi)^{-1/4}e^{\left(-\frac{1}{2}(X_n^-)^2\,-\, a_n^{+\dag}a_n^{-\dag} e^{2in\tau}\, + \,X^-_n a_n^{+\dag} e^{in\tau}\,+\,X^-_n a_n^{-\dag}e^{in\tau}\right)}|0\rangle\nonumber\\
=(2\pi)^{-1/4}e^{-\frac{1}{2}\left(X_n^--e^{in\tau}(a_n^{-\dag}+a_n^{+\dag})\right)^2}e^{\frac{1}{2}e^{2in\tau}\left((a_n^{+\dag})^2+(a_n^{-\dag})^2\right)}|0\rangle\hskip0,3cm
\qqq
where we have used 
\qq
a_n^-e^{a_n^{+\dag} f}|0\rangle=fe^{a_n^{+\dag}f}|0\rangle
\qqq
for $f$ a commuting operator with respect to $a^-_n$ and $a^{+\dag}_n$.
The $(D;n)$-Ishibashi state is  
\qq
|(D;n)\rangle=\prod_{n=1}^\infty\int \CD X^-_n |X^-_k\rangle\hskip3,5cm\nonumber\\
= \prod_{n=1}^\infty \exp\left(\frac{e^{2ni\tau}}{2}\left[(a_n^{+\dag})^2+(a_n^{-\dag})^2\right]\right)|0\rangle
\qqq
where we have performed a gaussian integration. The closed string is created at $\tau=0$ and annihilated at $\tau=\pi/T$.
The $\tau$ dependence can be factored out by integration
\qq
\prod_{n=1}^\infty\int_{-\infty}^0 d\sigma \exp\left(\frac{e^{2ni\tau}}{2}[(a_n^{+\dag})^2+(a_n^{-\dag})^2]\right)|0\rangle
\label{cdisk}
\qqq
where in the Appendix B we show that it gives
\qq
=\prod_{n=1}^\infty \int_{-\infty}^0 d\tau  e^{i\tau L_0^{(C;\ell)}}e^{\frac{1}{2}\left((a_n^{+\dag})^2+(a_n^{-\dag})^2\right)}|0\rangle\nonumber\\
=\,\int_{-\infty}^0 d\tau D_{ch}|H\rangle\hskip3,7cm
\qqq
and we see that the $H$-brane, described by the {\bf chiral} and {\bf squeezed} Ishibashi state 
\qq
|H\rangle=\prod_{n=1}^\infty e^{\frac{1}{2}\left((a_n^{+\dag})^2+(a_n^{-\dag})^2\right)}|0\rangle
\qqq
 couples naturally to the {\bf chiral closed string disk propagator}
\qq
D_{ch}=e^{i\tau L_0^{(C;\ell)}}
\qqq
with
\qq
L_0^{(C;\ell)}=\sum_{n=1}^\infty n(a_n^+)^\dag a_n^-\,+\,\sum_{n=1}^\infty n(a_n^-)^\dag a_n^+\hskip0,3cm .
\qqq  
In the Appendix C we show that the disk amplitude neglecting the zero-modes is 
\qq
\langle H|D_{ch}|H\rangle = \prod_{n=1}^\infty \frac{1}{1-z_n}\nonumber\\
z_n=e^{2in\tau}\hskip0,3cm .\hskip0,6cm
\qqq
Similar analysis follow for the case of closed $(N;n)$-boundary condition on the $X^+$ coordinate, where we only have to interchange left- with right-movers
\qq
|(N;n)\rangle= \prod_{n=1}^\infty \exp\left(\frac{e^{2ni\sigma}}{2}[(\tilde{a}_n^{+\dag})^2+(\tilde{a}_n^{-\dag})^2]\right)|0\rangle
\qqq
for $\sigma\sim \sigma+2\pi$. 
The vaccum is defined in the euclidian picture from the conditions that $T(z)|0\rangle$ and $\tilde T(\bar z)|0\rangle$ are well-defined as $z,\bar z\rightarrow 0$ wich implies
\qq
L_n|0\rangle =0 \hskip0.3cm,\hskip0,6cm \tilde L_n|0\rangle=0
\qqq
for all $n\geq -1$, and in particular is invariant under global conformal transformations with the $SL(2,\NR)_L\otimes SL(2,\NR)_R$ group generated by $L_{-1}$, $L_0$ and $L_1$ and their antiholomorphic counterparts. We will suppose that in the closed string channel, this definition is still valid in the Lorentz picture. Nevertheless a proper definition of the vacuum state in time-dependent conformal field theory is an open problem.

We note that 
\qq
\langle(D;n)|\tilde L^{(C,\ell)}_m|(D;n)\rangle=0\hskip2.8cm\nonumber\\
\langle(N;n)|L^{(C,\ell)}_m|(N;n)\rangle=0
\hskip1.5cm \forall m \in \NN
\qqq
since left- and right-chiral movers commute between each other.
If we want to interprete these conditions as $Diff(S^1)$ invariance, we must have $c=0$ and $\tilde c=0$ respectively.

For the fermionic contribution to the $H$-Ishibashi states, 
we add to the above states their Majorana components. In the closed string channel boundaries are set at $\tau=0,\,\pi/T$ and that gives conditions on the left-sector, see (\ref{mbc}). In the case at hand we have to flip left- from right-chiral mode sector when we move from the open string sector to the closed sector, see (\ref{LtoR}). In all we have the same conditions as in the open sector, $\Psi_R^-=0$. The non-vanishing counterpart is 
\qq
\Psi_L^-=\sum_{r\in\NZ+\nu}\psi^-e^{-ir\tau}
=\sum_{r\geq\nu}\theta_r^-
\qqq
where
\qq
\theta_r^-=\psi_r^-e^{-ir\tau}+\psi_r^{-\dag}e^{ir\tau}
\qqq
for $\tau=0,\,\,\pi/T$ and we have set $\psi_r^{-\dag}=\psi_{-r}^-$ for $r\geq\nu$. Again we set this condition as an eigenvalue for a given eigenstate $|\Theta^-\rangle$ 
\qq
\left(\theta_r^--\psi_r^-e^{-ir\tau}-\psi_r^{-\dag}e^{ir\tau}\right)|\Theta^-\rangle=0
\qqq
for $r>0$ and with solution
\qq
|\Theta^-\rangle=\prod_{r>0}\exp\left(-\frac{1}{2}(\theta_r^-)^2+\psi_r^{+\dag}\theta_r^-e^{ir\tau}+\psi_r^{-\dag}\theta_r^+e^{ir\tau}-\psi_r^{+\dag}\psi_r^{-\dag}e^{2ir\sigma}\right)|0\rangle\hskip0,1cm .
\qqq
Finally 
\qq
|N_\psi^-\rangle=\int \CD\Theta^-|\Theta^-\rangle=\prod_{r>0}\exp\left(\frac{1}{2}e^{2ir\tau}\psi_{-r}^+\psi_{-r}^+\right)
\qqq
for $\sigma=0,\,\pi/T$.


\chapter{Black hole quantum horizons}

\hskip0,5cm It was shown in the early seventies by Christodoulou \cite{christo},
Penrose and Floyd \cite{penrose} and Hawking \cite{hawking71}  
that, in a classical
theory, the area of a black hole horizon can  not decrease. 
 In his seminal paper \cite{bekenstein1}, Bekenstein used  this fact
  to postulate an identification between the horizon area and
 the black hole entropy.
 Later, Bardeen, Carter and Hawking found four laws of 
black hole mechanics analogous to the four
laws of thermodynamics \cite{hawking1}. The pioneer postulate of
Bekenstein was finally realized to be correct when
Hawking found, using quantum perturbations near the horizon, that
 a black hole radiates energy and behaves like a hot object
with a given temperature that is identified with the horizon surface gravity
\cite{hawking2}. 
We conclude that the four mechanical laws of black holes are in fact the 
four laws of thermodynamics. In particular we identify the area of the event horizon
 of {\it any} black hole with
 the Bekenstein-Hawking entropy (in Planck units) 
up to factor
\qq
S_{BH} = \frac{1}{4} A \hskip0,3cm .
\qqq
As in any thermodynamical system, the entropy must have some 
microscopic origin. Here the black hole entropy should be statistically 
given by
 a consistent quantum gravity theory, where the Ashtekar approach
is one candidate. Recently the case of the Schwarzschild 
black hole has been treated within this approach \cite{ashbh}. 

In \cite{bekenstein2} Bekenstein has done another important step to 
understand the black hole physics. There he suggested that the horizon area 
$A$ (and consequently the mass) must be quantized in Planck units.
This suggests that a quantum
horizon  behaves as a phase space formed by independent patches of 
equal Planck size areas, bearing 
in mind that each patch can locate one degree of freedom as in a usual
 phase space patchwork. This is now called the Holographic Principle
 \cite{thooft} \cite{susskind6}. 

\leavevmode\epsffile[-55 -40 200 520]{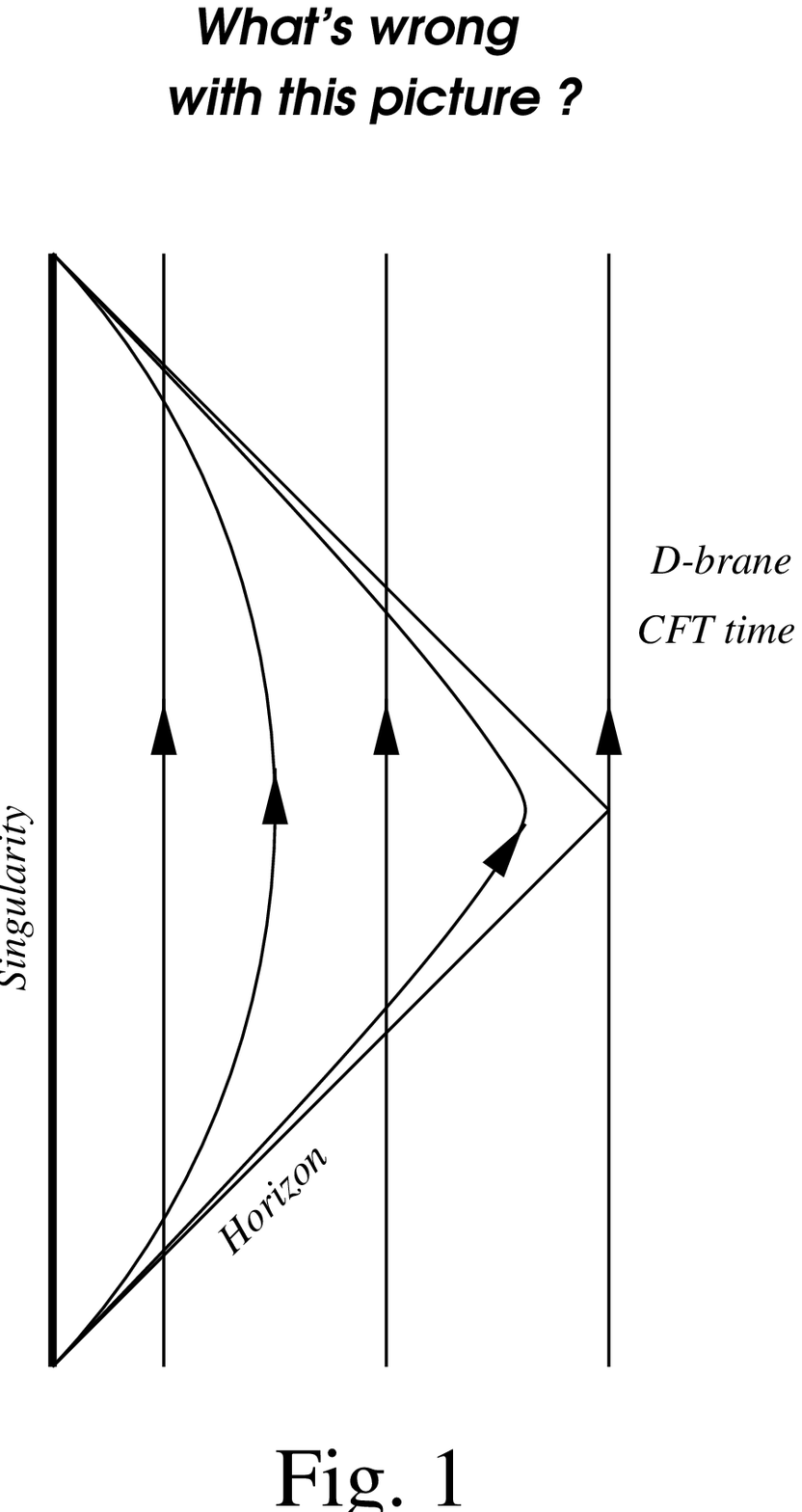}

The idea of a discrete spectrum proposed by Bekenstein was later
discussed  independently by Mukhanov \cite{muk} and Kogan
  \cite{kogan1}.  Arguments used in \cite{muk} were based on
entropy and further developments are  contained in \cite{bekmuk}. 
In recent years the discrete spectrum of quantum black holes was examined in
numerous papers, see for example \cite{vazwitten} and references therein. 
The black hole area quantization has been extensively discussed recently in
Ashtekar's approach to quantum gravity \cite{asht1}. However it is 
not clear whether the Ashtekar program is consistent with the Holographic Principle.

String theory is another candidate for a quantum gravity theory. 
In paper  \cite{kogan1} (see also \cite{kogan2}) a stringy approach to 
the black hole area quantization  was suggested. It was based on 
the consistency
 of a {\it chiral closed string} moving in a Euclidean black hole
 background. The fact that it leads to the same discrete spectrum as 
Bekenstein's was an indication that chiral sectors should be relevant to the 
statistical counting of black hole entropy. 

In \cite{strominger1} Strominger and Vafa have for the first time counted in a controlable 
way the correct number of microscopic states that gives rise to the Bekenstein-Hawking 
entropy of a extremal charged black hole, with a AdS geometry near the horizon, 
using D-brane technology. See \cite{maldacena} for a review and subsequent important 
developments. The counting of states is done in a dual theory of supergravity on AdS 
spacetime in the UV limit. That is the IR limit of a gauge theory living on the D-branes. 
Such duality has given rise to the AdS/CFT holographic principle \cite{maldacena2}.

Nevertheless, in the UV limit the black hole has become smaller then the string scale, 
the spacetime geometry becomes quite fuzzy and there is no concept of an horizon at all. 
Thus it is difficult to understand how one can explain the universality of 
the Bekenstein-Hawking entropy law \cite{strominger2}. 
Such a dual description is also problematic if we want to discuss the unitarity of 
the black hole evolution, see Fig. 1 for the embeding of the an extremal black 
hole geometry in a AdS spacetime \cite{strominger2}. Here curved rows represent 
the black hole time isometries and the straight lines the time isometries of the gauge theory that lives on the D-brane. From the AdS/CFT dual picture, straight lines describe a  unitary evolution of conformal field theory so by duality the black hole evolution should be unitary. However, it is hard to understand why this should be so: any initial configuration 
evolves into the black hole and we progressively loose information that was outside 
the black hole. Once inside it, the information cannot flow outside the black hole and 
continue to propagate freely to infinity. Clearly the evolution is not unitary! 
Nevertheless,  we stress here that the statistical count was performed by time 
independent String Theory and the information loss paradox is a time-dependent phenomena.

To summarize, in black hole physics we need a quantum model where the number of microstates 
gives rise in a controlable way to the Bekenstein-Hawking entropy and to an area 
quantization with the given discrete spectrum proposed by Bekenstein, consistent with 
the holographic principle. It is interesting to note that
 when we try to get a stringy description of a quantum 
horizon we either have to deal with open strings attached to $D$-branes \cite{strominger1}
 or with a chiral sector of closed strings \cite{kogan1}. The only common
feature is that both have {\it one} Virasoro
algebra. As we have seen, $H$-branes are naturally coupled to chiral closed strings. 
They are also associated with a non-commutativity along the light-cone coordinates, 
proposed by Susskind in \cite{susskindilp} to solve the information loss paradox. 
At last, from their time-dependent geometric shape, it is conceivable that $H$-branes 
can provide a {\it phenomenological picture} (instead of a {\it dual picture}) of 
the black hole horizon quantization. In this way it was conjectured in \cite{KR} 
that in a stringy picture, quantum horizons are described by chiral and non-normalized 
squeezed states.

In the following subsections, we review the BTZ black hole geometry and give some hints as to
how $H$-branes fit into it. We are, however, far from giving a clear picture on 
the BTZ black hole horizon quantization using $H$-branes.  


\section{The BTZ black hole}

\hskip0,5cm The BTZ black hole is a solution of the Einstein equations in 3-dimensions with a negative cosmological constant $\Lambda=-1/\ell^2$ \cite{btz}. The global geometry of this solution may be described by certain identifications on the $AdS_3$ spacetime. The solution depends 
on the ADM mass $M$ and angular momentum defined at spatial infinity $J$ that enter 
in the metric in the following way
\qq
ds^2=-(N^\perp)^2dt^2\,+\,f^{-2}dr^2\,+\,r^2(d\phi+N^\phi dt)^2
\qqq
with lapse and shift functions
\qq
N^\perp=f=\left(-M+\frac{r^2}{\ell^2}+\frac{J^2}{4r^2}\right)^{1/2}\,,\hskip1.0cm N^\phi=-\frac{J}{2r^2}\hskip0,3cm\nonumber\\
 \phi\sim\phi+2\pi\hskip0,3cm .
\qqq
There is a mass gap between the energy of the BTZ-solution with $M=J=0$ and the $AdS_3$ vaccum energy. The last is described by the BTZ-metric with $J=0$ and $M=-\frac{1}{8G}$. 
Naked singularities appear in the interpolations of those two solutions. Nevertheless, 
the existence of a tachyon does not mean instability of the $AdS_3$ vaccum. The $AdS$ 
vacuum is in fact perfectly stable.
When $M>0$ and $|J|\leq M\ell$, the solution describes a black hole with inner $r_-$ and outer $r_+$ horizon with
\qq
r_\pm^2=\frac{M\ell^2}{2}\left(1\pm\sqrt{1-(J/M\ell)^2}\right)
\qqq
i.e.,
\qq
M=\frac{r_+^2+r_-^2}{\ell^2}\,,\hskip1.0cm J=\frac{2r_-r_+}{\ell}\hskip0,3cm .
\qqq
The Bekenstein-Hawking entropy is given by one-quarter of the horizon area (in Planck units) 
\qq
S=\frac{A}{4G}=\frac{2\pi r^+}{4G}
\qqq
where $G$ is the 3-d Newton constant. The Hawking temperature is
\qq
T=\frac{r_+^2-r_-^2}{r_+\ell^2}\hskip0,3cm .
\qqq

In the euclidian picture, the $AdS$ geometry is a hyperbolic space whose geometry is given by the standart Poincare-metric
\qq
ds^2=\frac{\ell^2}{z^2}(dx^2+dy^2+dz^2)\,,\hskip1.0cm z>0
\qqq
where the cartesian coordinates may be expressed by the spherical coordinates 
\qq
(x,y,z)=(R\cos\theta\cos\chi,R\sin\theta\cos\chi,R\sin\chi)\hskip0,3cm .
\qqq
On the other hand, the outer region of the BTZ black hole is given by certain identifications on the above hyperbolic space. Explicity, we first do a coordinate transformation 
\qq
x=\left(\frac{r^2-r_+^2}{r^2-r_-^2}\right)^{1/2}\cosh\left(\frac{r_+}{\ell^2}-\frac{r_-}{\ell}\phi\right)\exp\left(\frac{r_+}{\ell}\phi-\frac{r_-}{\ell^2}t\right)\nonumber\\
y=\left(\frac{r^2-r_+^2}{r^2-r_-^2}\right)^{1/2}\sinh\left(\frac{r_+}{\ell^2}-\frac{r_-}{\ell}\phi\right)\exp\left(\frac{r_+}{\ell}\phi-\frac{r_-}{\ell^2}t\right)\nonumber\\
z=\left(\frac{r_+^2-r_-^2}{r^2-r_-^2}\right)^{1/2}\exp\left(\frac{r_+}{\ell}\phi-\frac{r_-}{\ell^2}t\right)
\qqq
followed be the discrete identifications in the sperical coordinates
\qq
(R,\theta,\chi)\sim(Re^{2\pi r_+/\ell},\theta+\frac{2\pi r_-}{\ell}, \chi)
\qqq
In particular, in the euclidian picture we see that for the non-extremal regime 
$r_+\neq r_-$, the event horizon shrinks to a circle ($x=y=0$) and at the extremal 
regime, it shrinks to a point
\qq
r=r_{\rm horizon}\rightarrow (x,y,z)=(0,0,0)\,,\hskip1.0cm (J=M\ell)
\qqq
where, as expected, the Hawking temperature vanishes.


\section{The $AdS_3$$/CFT_2$ holographic principle}

\hskip0,5cm Recently a correspondence between supergravity on $AdS_D$ and the Yang-Mills $\CN=4$ Theory 
at fixed point in one lower dimension was introduced \cite{maldacena2}. However, in 
a different context, gravity in $AdS_3$ with cosmological constant $\Lambda=-1/\ell^2$ 
was shown to be described at infinity by a CFT with central charge $c=\frac{3\ell}{2G}$ 
\cite{chd}. This can be realized by considering the following boundary conditions 
of an assymptotically $AdS_3$ metric
\qq
g_{tt}=-\frac{r^2}{\ell^2} + O(1)\hskip1.0cm
g_{t\phi} = O(1)\hskip1.0cm
g_{tr}=O(\frac{1}{r^3})\hskip1,0cm\nonumber\\
g_{rr}=\frac{\ell^2}{r^2}+O(\frac{1}{r^4})\hskip1.0cm
g_{r\phi}=O(\frac{1}{r^3})\hskip1.0cm
g_{\phi\phi}=r^2+O(1)\hskip0,3cm .
\qqq   
The diffeomorphism group that preserves these boundary conditions can be seen to be generated by the following vector fields $\zeta^a(r,t,\phi)$
\qq
\zeta^t=\ell(T^++T^-)+\frac{\ell^3}{2r^2}(\partial_+^2 T^+ + \partial_-^2 T^-)+O(\frac{1}{r^4})\nonumber\\
\zeta^\phi=T^+-T^--\frac{\ell^2}{2r^2}(\partial_+^2T^+-\partial_-^2T^-)+O(\frac{1}{r^4})\hskip0,5cm\nonumber\\
\zeta^r=-r(\partial_+T^++\partial_-T^-)+O(\frac{1}{r})\hskip3,0cm
\qqq 
where 
\qq
2\partial_\pm:=\ell\frac{\partial}{\partial t}\pm\frac{\partial}{\partial \phi}\hskip0,3cm .
\qqq
They are writen in term of two independent diffeomorphisms $T^\pm(r,t,\phi)=T^\pm(\frac{t}{\ell}\pm \phi)$. By expanding them in a Fourier series with $L_n\, (\tilde L_n)$ generating 
the diffeomorphism $T^\pm=e^{in(\frac{t}{\ell}\pm\phi)}$, it was shown that the 
asymptotic isometries are given by two commuting Virasoro algebras of central charge 
$c=\frac{3\ell}{2G}$. They represent a CFT that lives in the $(t,\phi)$ cylinder 
at spatial infinity.


\section{Counting the microstates of a BTZ black hole horizon}
\hskip0.5cm
{\it Strominger approach \cite{stro2}}: From the above $AdS_3/CFT_2$ duality, we define the ground state of the CFT, that lives in the cylinder at spatial infinity, by imposing the conditions on the zero-mode Virasoro operators $L_0=\tilde L_0=0$. We would like that by duality, the CFT vaccum state corresponds to the $J=M=0$ BTZ-solution. Furthermore, we may postulate the following holographic identifications between the Virasoro operators and the physical parameters that describe the spinning BTZ black hole
\qq
M=\frac{1}{\ell}(L_0+\tilde L_0)\nonumber\\
J=L_0-\tilde L_0\hskip0,3cm .
\qqq
In the semiclassical limit $\ell\gg G$ one may use the Cardy formula to count the asymptotic number of states in a CFT of a given central charge $c$
\qq
S=2\pi\sqrt{\frac{c\,L_0}{6}}+2\pi\sqrt{\frac{c\,\tilde L_0}{6}} 
\qqq
where in our case $c=\frac{3\ell}{2G}$.
By a short algebra we see that the result is in exact agreement with the Bekenstein-Hawking entropy. This argument should be valid in any consistent quantum theory of gravity, as is the case for String Theory. 

However the boundary microstates live on the cylinder at spatial infinity and it makes 
no direct connection to the number of states that we expected to find on the horizon, 
see \cite{Carlip} for a discussion of this point.
\vskip0.3cm
{\it Carlip approach}: A second way to count the microstates that gives rise to the 
Bekenstein-Hawking entropy may be found in \cite{Carlip2} where the author considers 
the horizon as a marginal trapped surface. It is known that 3-d gravity with local 
$AdS$ geometry may be treated in the bulk as a Chern-Simons theory. Treating the 
horizon as a boundary, the would-be-pure gauge degrees of freedom of the Chern-Simons 
theory become physical degrees of freedom of a WZW model at the 2-dimensional boundary 
surface with target group manifold $SL(2,R)_{-k}\otimes SL(2,R)_k$. Here the level is 
related to the cosmological constant by $k=\frac{\sqrt{2}\ell}{8G}$. Note that the 
CFT living at the horizon is different from the one that lives on the cylinder 
at spatial infinity. In particular, in the classical limit the present CFT has central 
charge $c=6$ whereas for the previous one $c\gg 1$.
 
It is known that in the WZW model, the Virasoro mode operators are formulated in terms 
of the chiral U(1)-current algebra $J$ and $\tilde J$ by the Sugawara construction. 
In particular $L_0$ and $\tilde L_0$ are 
\qq
L_0=-\frac{2}{2k+1}\sum_{m=-\infty}^\infty :J_{-n}^a J_n^b:\eta_{ab}\nonumber\\
\tilde L_0 =\frac{2}{2k-1}\sum_{m=-\infty}^\infty :\tilde J_{-n}^a \tilde J_n^b:\eta_{ab}\hskip0,3cm
\qqq
with the currents obeying the algebra
\qq
\left[J_m^a,\,\,J_n^b\right] = i\,f^{ab}_c\,J^c_{m+n}\,-\,km\eta^{ab}\delta_{m+n,0}\hskip0,4cm\nonumber\\
\left[\tilde J_m^a,\,\, \tilde J_n^b\right] = i\,f^{ab}_c\,\tilde J^c_{m+n}\,+\,km\eta^{ab}\delta_{m+n,0}\hskip0,3cm .
\qqq
Here we are using $(-,+,+)$ spacetime signature.
The vacuum state $|\Omega\rangle$ is annihilated by the current algebra generators $J_m$ and $\tilde J_m$ for $m>0$ and any physical state is given in terms of raising operators - $J_{-n}$ and $\tilde J_{-n}$ with $n>0$ - acting in the vaccum state 
\qq
|\phi,N,\tilde N\rangle \simeq |(n_1,a_1),(n_2,a_2),...\rangle|(\bar n_1,\bar a_1),(\bar n_2, \bar a_2),...\rangle\,=\nonumber\\=\,(J_{-n_1})^{a_1}
(J_{-n_2})^{a_2}...(\tilde J_{-\bar n_1})^{\bar a_1}(\tilde J_{-\bar n_2})^{\bar a_2}...|\Omega\rangle\hskip0,3cm.\hskip0,3cm
\qqq
Here $N=\sum a_in_i$ and $\tilde N=\sum \bar a_i\bar n_i$ are the number 
operators of the left and right moving oscillators respectively. The Hamiltonian describing the evolution of these states is, at the classical limit (large $k$), given by
\qq
H = L_0+\tilde L_0\,=\,N + \tilde N - \frac{2}{2k+1}(J_0)^2 + \frac{2}{2k-1}(\tilde J_0)^2\hskip0,3cm .
\qqq
The conjecture in \cite{Carlip} was to consider the horizon to have only left-moving excited states, i.e., $\tilde N=0$. Note that such chiral-state was postulated in \cite{kogan1} 
to describe a chiral closed string used to quantize the energy levels of the horizon. 

Further we impose the physical condition $H|\phi,N,0\rangle = 0$ where we have neglected 
the constant from normal ordering as it will not affect the final semi-classical result. 
Such a condition imposes a certain value for the number operator $N$.  Moreover from the boundary condition that was used to define the horizon as a trapped surface, the main 
contributions to the number of states came from the identifications
\qq
J_0|\Omega\rangle = -k\frac{r^+}{\sqrt 2 \ell}(2k+1)|\Omega\rangle\hskip0,5cm\nonumber\\
\tilde J_0|\Omega\rangle = -k\frac{r^+}{\sqrt 2 \ell}(2k-1)|\Omega\rangle\hskip0,3cm .
\qqq
It is interesting to see that the momentum operator $P=L_0-\tilde L_0$ is given by
\qq
P = N - 4k^3\frac{(r_+)^2}{\ell^2}\hskip0,3cm .
\qqq
Using the Cardy formula to count the number of states of the closed ``chiral excited'' string (remember that $\tilde N=0$)
\qq
S=2\pi\sqrt{\frac{cN}{6}}
\label{cardy}
\qqq
with $c=6$, it is easy to see that the condition $L_0+\tilde L_0=0$ translates now 
to $N=(\frac{r^+}{4G})^2$; when inserted in (\ref{cardy}) this gives 
the exact 
Bekenstein-Hawking entropy of the BTZ black hole.
Finally, the momentum operator takes the value
\qq
P=-\frac{2k^2 (r^+)^2}{\ell^2}(2k-1)\hskip0,3cm .
\qqq


\section{The black hole complementarity principle}

\hskip0,5cm The previous calculation by Strominger is consistent with the one performed by Callan 
and Maldacena for the $5$-dimensional extremal supersymmetric black hole, using the so 
called $D_1/D_5$ system, a gas of open strings stretching between a number $Q_5$ of
$D_5$-branes and $Q_1$ $D_1$-branes, all wrapped on a five-torus and for which we give 
the system a Kaluza-Klein momentum $N$ in one of the directions. In this picture, 
the entropy is given by 
\qq
Z=\prod_{n=1}^\infty \left(\frac{1+q^n}{1-q^n}\right)^{4Q_1Q_5} = \sum d(N)q^N
\qqq
where the integers $d(N)$ represent the degeneracy of the state with Kaluza-Klein momentum number $N$
\qq
d(N)\rightarrow\int dq\frac{Z(q)}{q^{N+1}}
\qqq
that can be estimated by the saddle point method. For $N\rightarrow \infty$, keeping 
fixed the product $Q_1Q_5$ fixed, this gives the entropy
\qq
S=\log d(N)\sim 2\pi\sqrt{NQ_1Q_5}
\qqq
and that agrees with the classical black hole entropy. The problem is that a similar model does not work for black holes whose Schwarzschild radius exceeds the compactification scale 
\cite{malsuss}. For these so called fat black holes, we take the limit where 
$Q_1$, $Q_5$ and $N$ tend to infinity in fixed proportion
\qq
S=\log d(N)\rightarrow N\log N
\qqq
which does not agree with the black hole entropy. 
The picture of the black hole complementarity comes from the D-brane picture in a 
straightforward way, by replacing the gas of $Q_1Q_5$ species by a single string and the 
level number $N$ by $N'=NQ_1Q_5$. The entropy of the fat black hole is the same as 
the entropy carried by a single long string with central charge $c=6$ (as in the  Carlip 
calculation) and a string tension $T\sim\frac{1}{g\alpha'Q_5}$.  
 
\leavevmode\epsffile[9 -10 150 170]{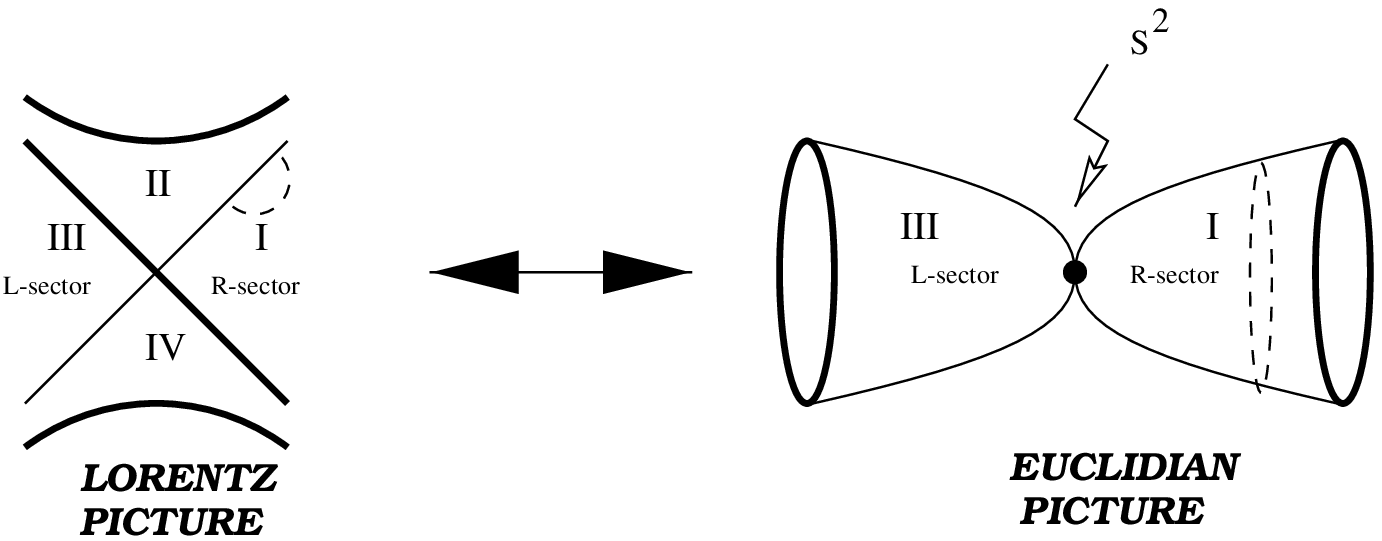}

The question we would like to address is to know exactly what are the properties of this long string. We suggest that such long string is open with the endpoints fixed at an $H$-brane. 
The states associated to the Bekenstein-Hawking entropy should be of open strings 
streching between stacks of $H$-branes. In the BTZ black hole, the state associated 
to this stack of $H$-branes should be chiral ($\tilde N=0$) with the conditions 
\qq
H^h|\phi,N,0\rangle = (\frac{r^+}{4G})^2\hskip1,5cm\nonumber\\
P^h|\phi,N,0\rangle = -(\frac{r^+}{4G})^2(2k-1)
\qqq
with 
\qq
H^h\,=\,\sum_{n>0}(\alpha_{-n}^+\alpha_n^-\,+\,\alpha_{-n}^-\alpha_n^+)\nonumber\\
P^h\,=\,\sum_{n>0}(\alpha_{-n}^+\alpha_n^-\,-\,\alpha_{-n}^-\alpha_n^+) 
\qqq
being the chiral hamiltionian and chiral momentum associated to null coordinates.
\vskip0.3cm
{\bf Summarizing:} We conjecture that quantum horizons - black hole, cosmological, etc - are described in String Theory by chiral and non-normalized squeezed states.


\chapter{Conclusion}

\hskip0,5cm In this work we note the absence in the known $D$-brane moduli space of Bosonic String 
Theory of an infinitelly boosted brane - a nullbrane. By $T$-duality, the nullbrane should 
also describe free open charged strings immersed in a critical constant 
electric field, where the pair-production of strings diverges. In both situations, 
from the worldsheet point of view the left- and right-chiral mode oscillators decouple 
 so that the theory becomes chiral and moreover the string endpoints are lightlike 
separate; the open string is naturally described by a worldsheet sigma model with 
null boundaries.\footnote{
 For an early discussion of boundary conditions of nullbranes in the context of conjugacy 
classes in group manifolds, see \cite{stanciu} (I
 thank V. Schomerus for calling attention to this reference).} 

 By working in the lightcone gauge, we show how the familiar worldsheet timelike 
boundaries are dynamically deformed to null boundaries at the critical limit. Null 
worldsheet boundaries set the correct boundary conditions that one would like to have 
on a nullbrane in Minkowskian spacetime - a Neumann condition for say 
$X^+$ and a Dirichlet condition for $X^-$, where $X^\pm$ are the target lightcone 
coordinates. We then carry the quantization of the system in the open string channel 
using the first-order formalism and find a space/time noncommutative geometry. 
From the analysis at the closed string channel, we calculate the $H$-brane Ishibashi 
states using the first-order formalism. Contrary to the case of $D$-branes that are 
described by non-chiral coherent Ishibashi states, $H$-brane Ishibashi states seem 
to be chiral and squeezed - $H$-branes are naturally coupled to chiral closed strings. 
Nevertheless, the result should be checked by carrying out a Cardy program in 
a time-dependent BCFT, but that is outside the scope of this M.Sc. thesis.

Let us note that $H$-branes where suggested by us from an initial attempt to find 
an alternative/complementary model in String Theory to the well established 
$AdS/CFT$ correspondence in order to describe quantum horizons. Because of their 
properties, we have conjectured that a stack of coinciding $H$-branes may 
{\it phenomenologically} describe any quantum horizon - black hole or cosmological - 
so that the states associated to quantum horizons should be in this way non-normalized, 
chiral and squeezed. 

We think that studying the concept of time in String Theory would answer open problems 
posed by quantum gravity, such as the information loss paradox. Nevertheless it is necessary 
to have at hand tools to describe String Theory in a mathematically consistent way, 
in particular for the study of branes - see \cite{GR} and references therein.

\chapter{Appendixes}

\nappendix{A}
\vskip0.5cm
Here we show that nullstrings described by the Schild action induce a spacetime noncommutativity on their endpoints. First note that the Schild action is already at first order
\qq
S_{\rm Schild}\,=\,\int \epsilon_{\mu\nu} dX^\mu\wedge dX^\nu
\qqq
interpreted as a tensionless open string immersed in a constant electromagnetic field $F_{\mu\nu}\propto \epsilon_{\mu\nu}$. We see that the equations of motion are given by $dF=0$ without involving the target coordinates and the boundary conditions are 
\qq
\epsilon_{\mu\nu}\delta X^\mu dX^\nu = 0\hskip0,3cm .
\qqq
The space of solutions is the set of coordinates with some constant value at the string endpoints
\qq
X^\mu(\tau,0)=x_0^\mu\hskip 2.0cm X^\mu(\tau,\pi)=x_1^\mu\hskip0,3cm .
\qqq
The symplectic structure of the phase-space is given by 
\qq
\Omega = \epsilon_{\mu\nu}\delta x_0^\mu\wedge\delta x_0^\nu - \epsilon_{\mu\nu}\delta x_1^\mu\wedge\delta x_1^\nu\hskip0,3cm .
\qqq
Here $\delta x_s^\mu$ ($s=0,1$) is a vector tagent to the space of solutions where the variation is given by a change of the constant $x_s^\mu$ from solution to solution. It is clear that it gives a noncommutative geometry at the nullstring endpoints. 


\nappendix{B}
\vskip0.5cm
In this appendix we calculate the chiral closed string disk propagator. First relabel the lightcone operators 
\qq
a_n^+\rightarrow a_n\hskip1.5cm (a_n^+)^\dag\rightarrow \tilde{a}_n^\dag\nonumber\\
a_n^-\rightarrow \tilde{a}_n\hskip1.5cm (a_n^-)^\dag\rightarrow a_n^\dag
\qqq
Consider
\qq
Z=hna_n^\dag a_n\hskip1.5cm Y=a_n^\dag a_n^\dag
\qqq
we have
\qq
[Z,Y]=2hnY
\qqq
so that by the Hausdorff formula we find 
\qq
e^Z e^Y=\exp(e^{2hn}Y)e^Z
\qqq
and since 
\qq
Z|0\rangle = 0
\qqq
we evaluate the integrand of (\ref{cdisk})  
\qq
\prod_{n=1}^\infty \exp\left[\frac{e^{2inx^+}}{2}\left((\tilde{a}_n^\dag)^2+(a_n^\dag)^2\right)\right]|0\rangle\nonumber\\
=\,e^{ix^+(L_0^{C}+\tilde{L}_0^{C})}\prod_{n=1}^\infty e^{\frac{1}{2}\left((\tilde{a}_n^\dag)^2+(a_n^\dag)^2\right)}|0\rangle
\qqq
where 
\qq
L_0^{(C)}=\sum_{n=1}^\infty na_n^\dag a_n\hskip1.5cm \tilde{L}_0^{(C)}=\sum_{n=1}^\infty n\tilde{a}_{n^\dag\tilde{a}_n}\hskip0,3cm .
\qqq
After relabel again to the lightcone oscillator modes
\qq
L_0^{(C)}+\tilde L_0^{(C)}\rightarrow L_0^{(C,\ell)}\,,
\qqq
 it is easy to see that (\ref{cdisk}) gives the desire result for the chiral closed string disk propagator. 


\nappendix{C}
\vskip0.5cm
In this appendix we calculate the chiral closed string amplitude between two H-branes. Consider
\qq
X=g(a_k)^2\hskip1.5cm Y=(a_k^\dag)^2
\qqq
and 
\qq
\langle 0|(a_k)^{2n}(a_k^\dag)^{2n}|0\rangle\,
=\,\langle 0|(a_k)^{2n-1}(a_k^\dag)^{2n-1}|0\rangle\,+\,\langle 0|(a_k)^{2n-1}a_k^\dag a_k(a_k^\dag)^{2n-1}|0\rangle\nonumber\\
=\,2n\langle 0|(a_k)^{2n-1}(a_k^\dag)^{2n-1}|0\rangle\hskip4,85cm\nonumber\\
=\,(2n)!\hskip8,25cm
\qqq
so that 
\qq
\langle 0 |X^n e^Y|0\rangle = \frac{1}{n!}\langle 0|X^n Y^n|0\rangle = \frac{g^n}{n!}(2n)!
\qqq
and finally
\qq
\langle 0|e^X e^Y|0\rangle = \sum_{n=0}^\infty \frac{1}{n!}\langle 0|X^n e^Y|0\rangle\nonumber\\
= \sum_{n=0}^\infty g^n\frac{(2n)!}{n!n!}\hskip1,05cm\nonumber\\
= \sum_{n=0}^\infty C_n^{2n} g^n\hskip1,35cm\nonumber\\
= \frac{1}{\sqrt{1-4g}}\hskip0,3cm . \hskip1,05cm
\qqq
The disk amplitude  presented in the text follows easily. We have to relabel the previous oscillators as the lightcone oscillators.


\end{document}